\newcommand*{\ATLASLATEXPATH}{latex/}
\documentclass[cernpreprint,texlive=2011,txfonts,UKenglish]{\ATLASLATEXPATH /atlasdoc}

\usepackage[subfigure=true]{\ATLASLATEXPATH atlaspackage}
\usepackage{\ATLASLATEXPATH atlasbiblatex}
\usepackage{\ATLASLATEXPATH atlascontribute}
\usepackage{\ATLASLATEXPATH atlasphysics}

\graphicspath{{logos/}{figures/}{logos/}}

\usepackage{MinBias13TeVPaper-defs}

\hypersetup{pdftitle={ATLAS Minbias},pdfauthor={The ATLAS Collaboration}}

\AtlasTitle{Charged-particle distributions in $\sqrt{s}=13$~TeV $pp$ interactions measured with the ATLAS detector at the LHC
}

\author{The ATLAS Collaboration}
\date{\today}

\AtlasRefCode{STDM-2015-02}
\AtlasNote{ATL-COM-PHYS-2015-346}
\PreprintIdNumber{CERN-EP-2016-014}

\AtlasDate{\today}

\AtlasJournal{Phys.\ Lett.\ B.}
\AtlasJournalRef{\PLB 758 (2016) 67}
\AtlasDOI{10.1016/j.physletb.2016.04.050}

\AtlasAbstract{%
Charged-particle distributions are measured in
 proton--proton collisions at a
centre-of-mass energy of 13~TeV, using a data sample of nearly 9
million events, corresponding to an integrated luminosity of 170~$\mu$b$^{-1}$, recorded by the ATLAS
detector during a special Large Hadron Collider fill.
The charged-particle multiplicity, its dependence on transverse
momentum and pseudorapidity and the dependence of the mean
transverse momentum on the charged-particle multiplicity are presented.
The measurements are performed with charged particles with transverse
momentum greater than 500~MeV and absolute pseudorapidity
less than 2.5, in events with at least one charged particle
satisfying these kinematic requirements.
Additional measurements in a reduced phase space with absolute pseudorapidity
less than 0.8 are also presented, in order to compare with other experiments.
The results are corrected for detector effects, presented as
particle-level distributions and are compared to the predictions of various Monte Carlo event generators.

}

\AtlasCoverSupportingNote{Charge particle multiplicity in pp interactions at 13 TeV measured with the ATLAS detector}{https://cds.cern.ch/record/2011604}

\AtlasCoverCommentsDeadline{25 January 2016}

\AtlasCoverAnalysisTeam{Anthony Morley (Sydney), Thorsten Kuhl (DESY), Geraldine Conti (CERN), Emily Nurse (London UC), Krzysztof Wozniak (Cracow), Stewart Martin-Haugh (RAL), Olivier Arnaez (CERN), Douglas Michael Schaefer (CERN), Jiri Kvita (Olomouc), Edvin Sidebo (Stockholm KTH), Jan Thomas Kuechler (Wuppertal), Daniela Boerner (Wuppertal), Oldrich Kepka  (Prague AS), Filip Nechansky  (Prague AS), Matous Vozak  (Prague AS), Maria Roberta Devesa (Buenos Aires), Roberto Iuppa (Roma II), Roman Lysak (Prague AS), Wolfgang Lukas (Innsbruck), Christoph Eckardt (DESY), Valentina Cairo (Cosenza), Matthias Danninger (Vancouver UBC), Tulay Cuhadar Donszelmann (Sheffield), Andrew Chisholm (Birmingham), Yuri Kulchitsky (Minsk AC), Elena Plotnikova (JINR Dubna), Hideyuki Oide (CERN), Alison Lister ((Vancouver UBC), Simone Pagan Griso (Berkeley), Heather Gray (CERN), Deepak Kar (Witwatersrand), Andreas Salzburger (CERN), Tim Martin (Warwick), Pavel Tsiareshka (JINR)}

\AtlasCoverEdBoardMember{Patricia Ward~(chair), Alejandro Alonso, Peter Maettig, Marjorie Shapiro}

\AtlasCoverEgroupEditors{atlas-STDM-2015-02-editors@cern.ch}

\AtlasCoverEgroupEdBoard{atlas-STDM-2015-02-editorial-board@cern.ch}

\begin{document}

\maketitle

\section{Introduction}
\label{sec:intro}
Charged-particle measurements in proton--proton ($pp$) collisions provide insight into the strong interaction in the
low-energy, non-perturbative region of quantum chromodynamics (QCD). 
Particle interactions at
these energy scales are typically described by
QCD-inspired models implemented in Monte Carlo (MC) event generators  with free
parameters that can be constrained by such measurements.
An accurate description of low-energy strong interaction processes is essential for simulating 
single $pp$ interactions as well as the effects of multiple $pp$ interactions at high instantaneous luminosity in hadron colliders.
Charged-particle distributions have been measured previously
in $pp$ and proton--antiproton collisions at various
centre-of-mass energies
~\cite{STDM-2010-06, CMS-QCD-10-004,
  cmsminbias_2,CMS-QCD-09-010,alice_mb2,Aaltonen:2009ne,Khachatryan:2015jna}
(and references therein).

This paper presents inclusive measurements of primary charged-particle
distributions in $pp$ collisions at a centre-of-mass energy of $\sqrt{s}=13$~TeV,
using data recorded by the ATLAS experiment~\cite{PERF-2007-01}  at
the Large Hadron Collider (LHC)
corresponding to an integrated luminosity of approximately
170~$\mu$b$^{-1}$.
Here inclusive means that all processes in $pp$ interactions are
included and no attempt to correct for certain types of process, such
as diffraction, is made.
These measurements, together with previous results, shed light
on the evolution of charged-particle multiplicities with
centre-of-mass energy, which is poorly constrained.
A strategy similar to that in Ref.~\cite{STDM-2010-06} is used,
where more details of the analysis techniques are given.
 The
distributions are measured using tracks from primary charged particles,
corrected for detector effects, and are presented as
inclusive distributions in a well-defined kinematic region. 
Primary charged particles are defined as charged particles with a mean
lifetime $\tau > 300$~ps, either directly produced in $pp$
interactions or from subsequent decays of directly produced particles with $\tau < 30$~ps;
particles produced from decays of particles with $\tau > 30$~ps, called secondary particles,
are excluded. This
   definition differs from earlier analyses in that charged particles with  a
 mean lifetime $30 < \tau < 300$~ps were previously
 included. These are charged strange baryons and have been removed due
 to the low efficiency of reconstructing them.\footnote{Since strange
   baryons tend to decay within the detector volume, especially if
   they have low momentum, they often do not leave enough hits to
   reconstruct a track, leading to a track reconstruction  efficiency
   of approximately 0.3\%.}
All primary charged particles are required to have a  
momentum component transverse to the beam direction,\footnote{ATLAS uses a right-handed coordinate system with its origin at the nominal interaction point (IP) in the centre of the detector and the $z$-axis along the beam pipe. The $x$-axis points from the IP to the centre of the LHC ring, and the $y$-axis points upward. Cylindrical coordinates $(r,\phi)$ are used in the transverse plane, $\phi$ being the azimuthal angle around the beam pipe. The pseudorapidity is defined in terms of the polar angle $\theta$ as $\eta=-\ln\tan(\theta/2)$.}  
\pta, of at least 500~MeV and absolute pseudorapidity, $|\eta|$, less than 2.5. Each event is required to have at least one primary charged particle. 

In these events the following distributions are measured: 
	$$
	\dndeta,   \ \
	\dndpt,   \ \
	{\mathrm{and}} \ \ 
	\dNdn   \ \
	$$
as well as the mean \pta\ (\meanpt{}) of all primary charged particles versus \nch.
Here \nch\ is the number of primary charged particles in an event, 
\nev\ is the number of events with $\nch \geq 1$, and \Nch\ is the
total number of primary charged particles in the data sample.\footnote{The factor $2 \pi  p_\mathrm{T}$ in the \pta\ spectrum comes from the Lorentz-invariant definition of the cross section in terms of $\mathrm{d}^3p$. 
The results could thus be interpreted as the massless approximation to $\mathrm{d}^3p$. }
The measurements are also presented in a phase space that is common to
the ATLAS, CMS~\cite{1748-0221-3-08-S08004} and
ALICE~\cite{1748-0221-3-08-S08002} detectors in order to ease
comparison between experiments. For this purpose an additional
requirement of $|\eta| < 0.8$ is made for all primary charged particles. These
results are presented in Appendix~\ref{sec:reduced}.
Finally, the mean number of primary charged particles for $\eta=0$ is
compared to previous
measurements at different centre-of-mass energies.
The measurements are compared to particle-level MC predictions.

The remainder of this paper is laid out as follows. The relevant
components of the ATLAS detector are described in
Section~\ref{sec:detector}. The MC event generators and
detector simulation used in the analysis are introduced in
Section~\ref{sec:mc}. The selection criteria applied to the data and
the contributions from background events are discussed in
Sections~\ref{sec:evt_sel} and~\ref{sec:background}
respectively. The selection efficiency and corresponding
corrections to the data are discussed in Sections~\ref{sec:eff-corr}
and~\ref{sec:corr} respectively.
The corrected results are compared to theoretical
predictions in Section~\ref{sec:result} and a conclusion is given in Section~\ref{sec:conclusion}.
The measurement of primary charged particles in the reduced phase space of $|\eta| < 0.8$ is presented in Appendix~\ref{sec:reduced}.
\section{ATLAS detector}
\label{sec:detector}
The ATLAS detector 
covers  almost the whole solid angle
around the collision point with 
layers of tracking detectors, calorimeters and muon chambers. 
For the measurements presented in this
paper,  the tracking devices and the trigger system are of particular importance.

The inner detector (ID) has full coverage in $\phi$ and covers the
pseudorapidity range $|\eta|~<~2.5$.
It consists of a silicon pixel detector (pixel), a silicon microstrip
detector (SCT) and a transition radiation straw-tube tracker
(TRT). 
These detectors span a sensitive radial distance from the
interaction point of
33--150~mm, 299--560~mm and 563--1066~mm respectively, and are situated inside
a solenoid that provides a 2~T  axial magnetic field.
The  barrel (each end-cap) consists of four (three) pixel
layers, four (nine) double-layers of single-sided silicon microstrips
with a 40~mrad stereo
angle between the inner and outer part of a double-layer, and 73 (160) layers of TRT straws.
The innermost pixel layer, the insertable B-layer
(IBL)~\cite{BLayerTDR}, was added between Run 1 and Run 2 of the LHC,
around a new narrower (radius of 25~mm) and thinner beam pipe. It is
composed of 14 lightweight staves arranged in a cylindrical geometry, each made of 12 silicon planar sensors in its central region and 2 $\times$ 4 3D sensors at the ends.  
The IBL pixel dimensions are $50 \times 250\,\mathrm{\mu m}^2$  in
the $\phi$ and $z$ directions (compared with $50 \times 400\,\mathrm{\mu
  m}^2$ for other pixel layers).
The smaller radius and the reduced pixel size result in improvements of both the transverse and longitudinal impact parameter resolutions.
 In addition, new services 
have been implemented which significantly reduce the material at the boundaries of the active tracking volume. 
A track from a charged particle traversing the barrel detector 
typically has 12
silicon measurement points (hits), of which four are pixel and eight SCT, and more than
30 TRT straw hits.

The ATLAS detector employs a two-level trigger system: the level-1 hardware stage
(L1) and the high-level trigger software stage (HLT).
This measurement uses the L1 decision from the minimum-bias trigger
scintillators (MBTS), which were replaced between Run 1 and Run 2. The MBTS are mounted at each end of the detector in front
of the liquid-argon end-cap calorimeter cryostats at $z = \pm 3.56$~m and
segmented into two rings in pseudorapidity ($2.07 < |\eta|~<~2.76$ and $2.76 <
|\eta|~<~3.86$). The inner ring is segmented into eight azimuthal sectors while
the outer ring is segmented into four azimuthal sectors, giving a total of
twelve sectors per side. The MBTS trigger selection
used for this paper requires one counter above threshold from either side of
the detector and is referred to as a single-arm trigger. The efficiency of this
trigger is studied with an independent control trigger. The control trigger
selects events randomly at L1 which are then filtered at HLT by requiring at
least one reconstructed track with $\pta > 200$~MeV.

\section{Monte Carlo event generator simulation}\label{sec:mc}
The \py~\cite{Sjostrand:2007gs},
\epos~\cite{Porteboeuf:2010um} and \qgsjet~\cite{Qgsjet}
MC generators are used to correct the data for detector effects and to compare with 
particle-level corrected data.
A brief introduction to the relevant parts of these
event generators is given below.

In \py\, inclusive hadron--hadron interactions are described by a model that
 splits the total inelastic cross section into non-diffractive (ND)
processes, dominated by $t$-channel gluon exchange, and diffractive
processes involving a colour-singlet exchange.  The simulation of ND
processes includes multiple parton--parton interactions (MPI).
The diffractive processes are further divided into
single-diffractive dissociation (SD), where one of the initial protons remains
intact and the other is diffractively excited and dissociates, and
double-diffractive dissociation (DD) where both protons dissociate.
The sample contains approximately 22\%\ SD and 12\%\ DD processes.
Such events tend to have large gaps in particle production at central
rapidity.
A pomeron-based approach is used to
describe these events ~\cite{corke_interleaved_2011}.

\epos\ provides an implementation of a parton-based
Gribov--Regge~\cite{Drescher:2000ha} theory, which is an effective QCD-inspired
 field theory
describing hard and soft scattering simultaneously. 

\qgsjet\ provides a phenomenological treatment of hadronic and nuclear
interactions in the Reggeon field theory framework ~\cite{Reggeon}. The soft and semi-hard parton
processes are included in the model within the ``semi-hard pomeron'' approach.
\epos\ and \qgsjet\ calculations do not rely on the standard parton
distribution functions (PDFs) as used
in generators such as \py.

Different settings of model parameters optimised to
reproduce existing experimental data are used in the simulation. These settings are referred to as tunes. 
For \py\ two tunes are used, \atwo~\cite{ATL-PHYS-PUB-2011-014} and \monash~\cite{Skands:2014pea}; for \epos\
the \lhc~\cite{Pierog:2013ria} tune is used. 
\qgsjet\ uses the default tune from the generator.
Each tune utilises 7~TeV minimum-bias data and is
summarised in Table~\ref{tab:tunes-compare}, together with the version
of each generator used to produce the samples.
\begin{table}
\begin{center}
	\caption{Summary of MC tunes used to compare to the corrected
          data. The generator and its version are given in the first
          two columns, the tune name and the PDF used are given in the
          next two columns.}
	\begin{tabular}{lllll} 
          \hline 
          Generator     & Version             & Tune                        & PDF                           \\
          \hline  
          \py\             & 8.185                &  \atwo\ & \mstwlo~\cite{Martin:2009iq}    \\
          \py\             & 8.186               & \monash\    & \nnpdflo~\cite{Ball:2012cx}          \\
          \epos\         &  LHCv3400            & \lhc\
          & N/A                           \\ 
          \qgsjet\     & \textsc{II-04}   & default & N/A                       \\ 
         \hline 
	\end{tabular}
        \label{tab:tunes-compare}
\end{center}
\end{table}
The \py\ \atwo\ sample, combined with a single-particle MC simulation
used to populate the high-\pta{} region, is used to derive the detector corrections
for these measurements.
All the events are processed through the ATLAS detector simulation
program~\cite{SOFT-2010-01}, which is based on {\textsc{geant}}{\scriptsize
  4}~\cite{Agostinelli:2002hh}. They are then reconstructed and analysed by the same program chain used for the data.

\section{Data selection}\label{sec:evt_sel}
The data were recorded during a period with a special configuration of
the LHC with low beam currents and reduced beam focusing, and thus giving a low expected mean
number of interactions per bunch crossing, $\avgmu=0.005$. 
Events were selected from colliding proton bunches using a trigger which required one or more MBTS counters above threshold on either side of the detector.  

Each event is required to contain a primary vertex, 
reconstructed from at least two tracks with a minimum \pta\ of 100~MeV, as described in Ref.~\cite{Piacquadio:2008zzb}. To
reduce contamination from events with more than one interaction in a 
bunch crossing, events with a second vertex containing four or more
tracks are removed. 
Events where the second vertex has fewer than four tracks are not removed. These are dominated by contributions where a secondary interaction is reconstructed as another primary vertex or where the primary vertex is split into two vertices, one with few tracks.
The fraction of events rejected by the veto on additional vertices due to
split vertices or secondary interactions is
estimated in the simulation to be 0.02\%, which is negligible and
therefore ignored.

Track candidates are reconstructed~\cite{newt1,newt2} in the silicon detectors and 
then extrapolated to include measurements in the TRT.
Events are required to contain at least one selected track,
passing  the following criteria:
$\pta>500$~MeV and  $|\eta | < 2.5$; at least one pixel hit and at least
six SCT hits, with the additional requirement of an innermost-pixel-layer  hit if expected\footnote{A hit is expected if the
  extrapolated track crosses an active region of a pixel module that
  has not been disabled.} (if a hit in the innermost layer is not expected, the next-to-innermost hit
is required if expected); $| \dzerobs |< 1.5$~mm,  where  the transverse impact parameter,
\dzerobs, is calculated with respect to the measured beam line position; and
$|\zzsint| < 1.5$~mm, where $\zBL$
is the difference
between the longitudinal position of the track along the beam line at the
point where $\dzerobs$ is measured and the longitudinal position of the primary
vertex, and $\theta$ is the polar angle of the track.
Finally, in order to remove tracks
with mismeasured \pta\ due to interactions with the material or other effects,
the track-fit $\chi^2$ probability
is required to be greater than $0.01$ for tracks with $\pta > 10$~GeV.
There are 8.87 million events selected, containing a total of 106 million
selected tracks.

\begin{figure*}
	\begin{center}
	\subfigure[]{
	\includegraphics[width=0.48\textwidth]{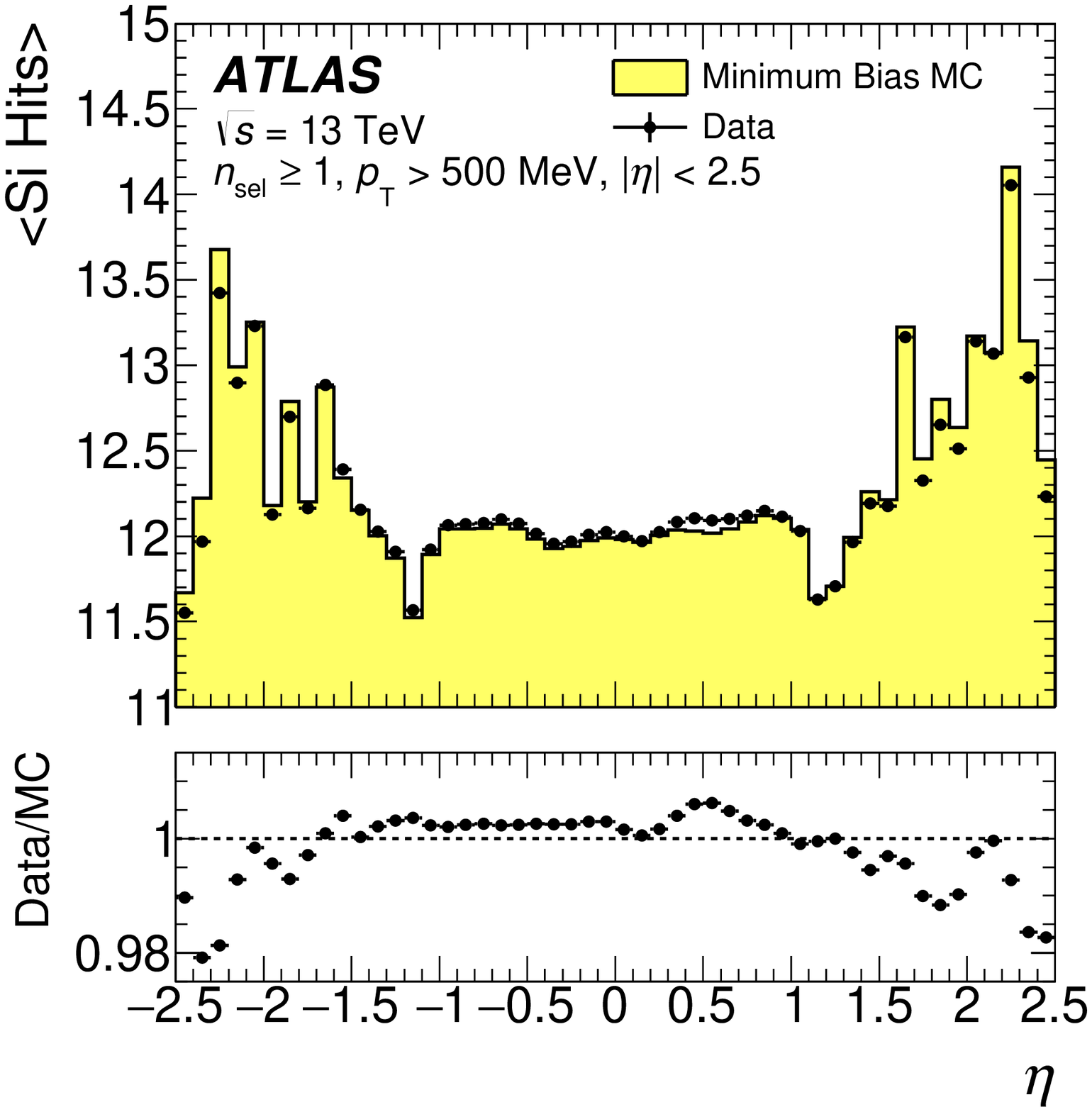}
                \label{fig:track-si}
	}
	\subfigure[]{
	\includegraphics[width=0.48\textwidth]{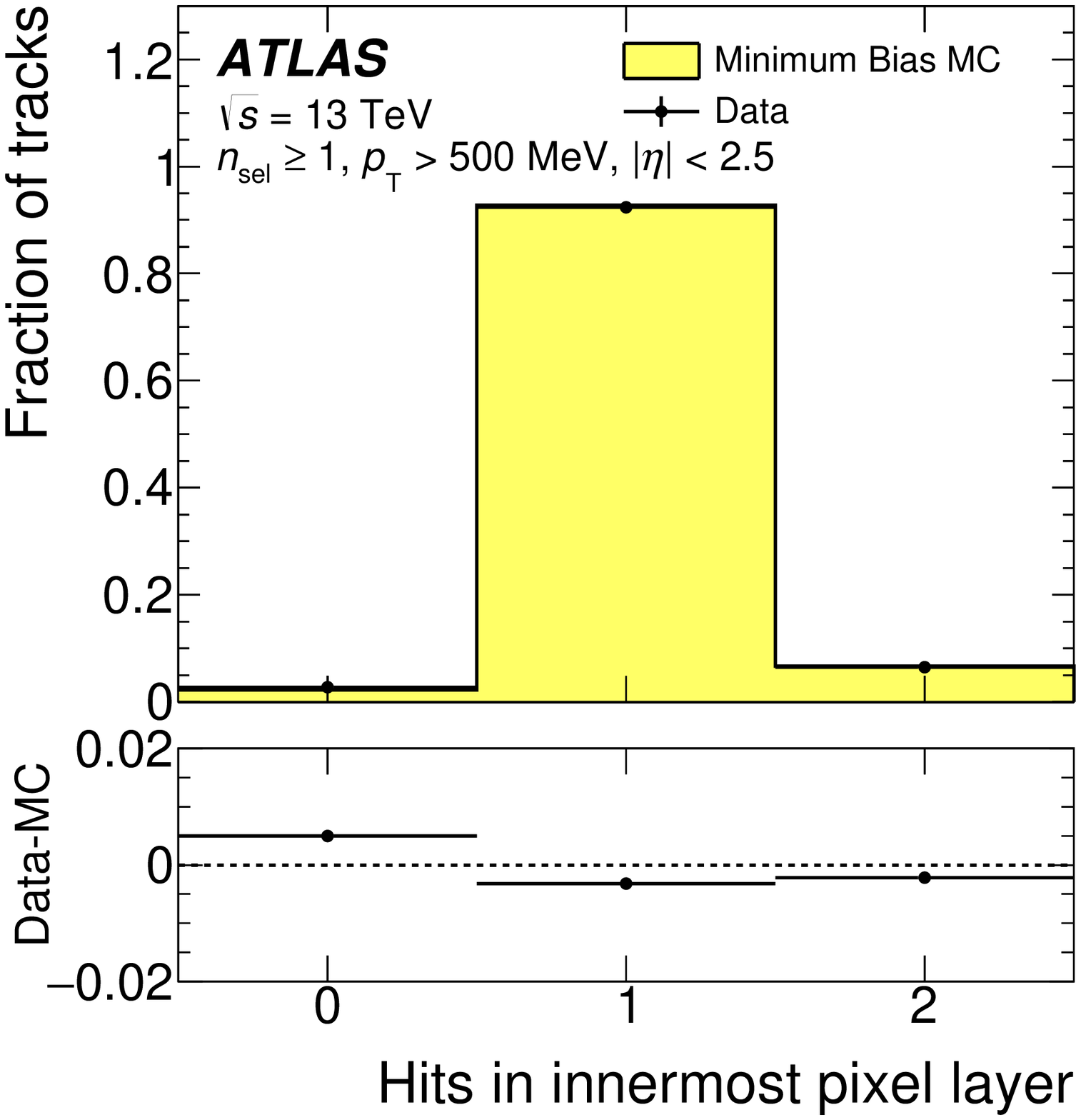}
                \label{fig:track-IBL}
	}
	\subfigure[]{
   	  \includegraphics[width=0.48\textwidth]{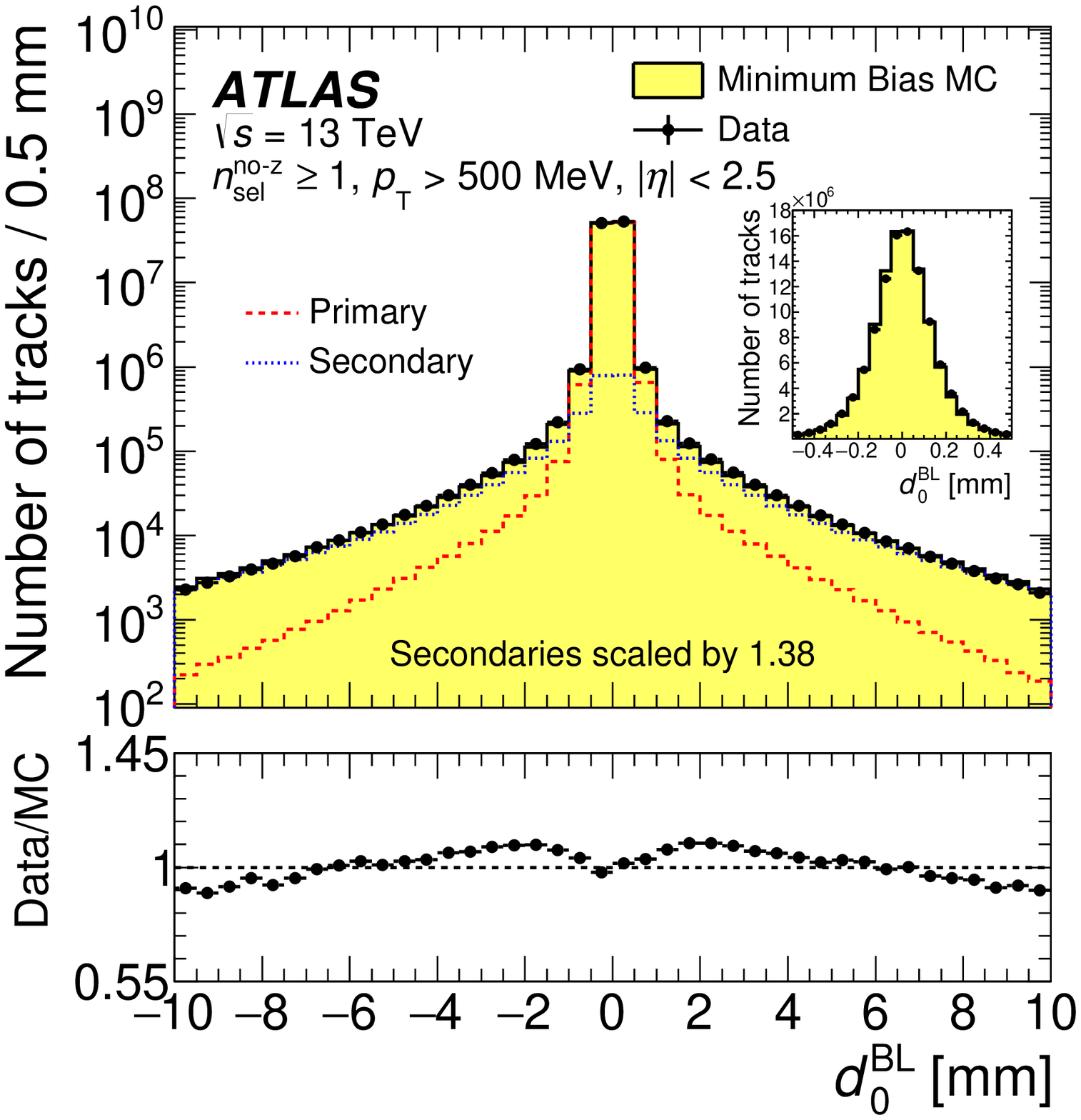}          
       \label{fig:track-d0}
	}
	\subfigure[]{
	\includegraphics[width=0.48\textwidth]{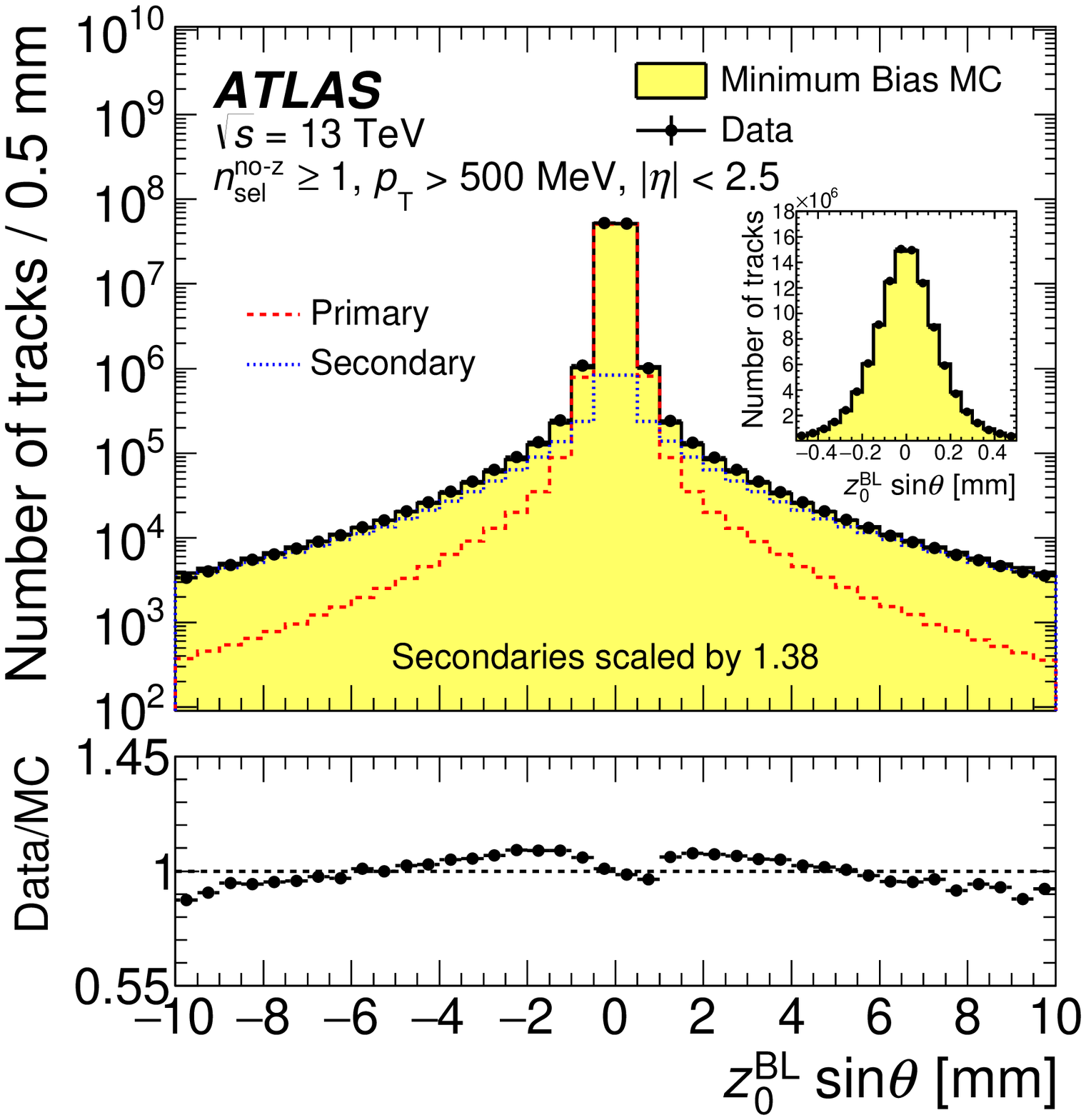}
               \label{fig:track-z0}
	}
	\end{center}
	\caption {Comparison between data and \py{} \atwo{} simulation for (a) the average number of
          silicon hits per track, before the requirement on the number of SCT hits is applied, as a
          function of pseudorapidity, $\eta$; (b) the number of
          innermost-pixel-layer hits on a track before the requirement on 
          the number of innermost-pixel-layer hits is applied; (c) the transverse
          impact parameter distribution of the tracks, prior to any requirement on the transverse
          impact parameter, calculated with
          respect to the average beam position, \dzerobs{}; and (d)
          the difference between the longitudinal position of the track along the beam line at the
          point where \dzerobs{} is measured and the longitudinal
          position of the primary vertex projected to the plane transverse to
the track direction, \zzsint, prior to any requirement on \zzsint.
          The uncertainties are the statistical uncertainties of the data.
          In (c) and (d) the separate contributions from tracks coming from primary and secondary
           particles are also shown and the fraction of secondary particles in the simulation is
           scaled by 1.38 to match that seen in the data, with the final simulation distributions 
           normalised to the number of tracks in the data.
           The  inserts in the panels for (c) and (d) show the distributions on a linear scale.
        }
  \label{fig:track_val}
\end{figure*}

The performance of the ID track reconstruction in the 13~TeV data and its simulation is studied in
Ref.~\cite{ATL-PHYS-PUB-2015-018}. Overall,  good agreement between data and
simulation is observed. 
Figure~\ref{fig:track_val} shows selected performance plots particularly
relevant to this analysis.
Figure~\ref{fig:track-si}  shows the average number of silicon hits as
a function of $\eta$. There is reasonably good agreement, although
discrepancies of up to 2\%\ (in the end-caps) are seen; however, these 
have a small effect on the track reconstruction efficiency. 
The discrepancies are due to differences between data and simulation in
the number of operational detector elements and an imperfect description of
the amount of detector material between the pixel detector and the SCT. 
The impact on the results of these discrepancies 
is discussed in Section~\ref{sec:trkeff}.
Figure~\ref{fig:track-IBL}  shows the fraction of tracks with a given number of IBL hits per
track. There is a difference of 0.5\%\ between data and simulation in the fraction of tracks with zero IBL
hits, coming predominantly from a difference in the rate of tracks
from secondary particles, which is discussed in more detail in Section~\ref{sec:background}.
 A systematic
uncertainty due to the small remaining difference in the efficiency of the
requirement of at least one IBL hit is discussed in Section~\ref{sec:eff-corr}.
Figures~\ref{fig:track-d0} and~\ref{fig:track-z0} show the \dzerobs{}
and \zzsint{} distributions respectively. 
In these figures the fraction of tracks from secondary particles in
simulation is 
scaled to match the fraction seen in data, and the separate contributions from
tracks from primary and secondary particles are shown. 
This, along with the differences between simulation and data, which have
a negligible impact on the analysis, are discussed in Section~\ref{sec:background}.

\section{Background contributions and non-primary tracks}
\label{sec:background}

The contribution from non-collision background events, such as proton 
interactions with residual gas molecules in the beam pipe, is
estimated using events that pass the full event selection but occur when only one of the
two beams is present.
After normalising to the contribution expected in the selected data
sample (using the difference in the time of the MBTS hits on each side
of the detector, which is possible as background events with hits on
only one side are negligible) a contribution of less than 0.01\%\ of events is
found from this source, which is
negligible and therefore neglected.
Background events from cosmic rays, estimated by
considering the expected rate of cosmic-ray events compared to the
event readout rate, are also found to be negligible and therefore neglected.

The majority of events with more than one interaction in the same
bunch crossing are removed by the rejection of events with more than
one primary vertex. Some events may survive because 
the interactions are very close in $z$ and are merged together. 
The probability to merge vertices is estimated by inspecting 
the distribution of the difference in the $z$ position of pairs
of vertices ($\Delta z$).  This distribution displays a deficit around $\Delta z=0$ due to vertex merging.
The magnitude of this effect is used to estimate the probability of merging vertices, which is 3.2\%. 
 When this is combined with the number  of expected additional interactions for $\avgmu=0.005$, the remaining
 contribution from tracks from additional interactions 
is found to be less than 0.01\%, which is
negligible and therefore neglected. The additional tracks in events
 in which the second vertex has fewer than four
associated tracks are mostly rejected by the $\zzsint$ requirement,
and the remaining contribution is also negligible and neglected.

The contribution from tracks originating from secondary particles is
subtracted from the number of reconstructed tracks before correcting
for other detector effects. These particles are due to hadronic interactions, photon
conversions and decays of long-lived particles. There is also a
contribution of less than 0.1\%\ from fake tracks (those formed by a
random combination of hits or from a combination of hits from several
particles); these are neglected.
The contribution of  tracks from secondary particles is estimated using simulation predictions for the shapes of
the \dzerobs\ distributions for tracks from primary and secondary particles satisfying  all
track selection criteria except the one on \dzerobs. 
These predictions form templates that are fit to the data in order to
extract the relative contribution of tracks from secondary particles.
The Gaussian core of the distribution is dominated by the tracks from
primary particles, with a width determined by their \dzerobs\ resolution; 
tracks from secondary particles dominate the tails. 
The fit is
performed in the region $4 < |\dzerobs| < 9.5$~mm, in order to reduce the
dependence on the description of the $\dzerobs$ resolution, which
affects the core of the distribution. From the fit, it was determined that the 
fraction of tracks from secondary particles in simulation needs to be scaled by
a factor $1.38 \pm  0.14$. This indicates that 
$\left( 2.3 \pm  0.6 \right)\%$ of  tracks satisfying the final track
selection criteria ($|\dzerobs| < 1.5$~mm) originate from secondary particles, where systematic uncertainties
are dominant and are discussed below. Of these tracks 6\%{} come from
photon conversions and the rest from hadronic interactions or long-lived decays.
The description of the $\eta$ and \pta\ dependence of this contribution 
is modelled sufficiently accurately by the simulation that no additional correction is required.    
Figure~\ref{fig:track-d0} shows the \dzerobs\ distribution for data
compared to the simulation with the fraction of tracks from secondary particles  scaled to
the fitted value. 
A small disagreement is observed in the core of the \dzerobs\ distribution.
This has no impact in the tail of the distribution used for the fit.
The dominant systematic uncertainty stems from the 
interpolation of the number of tracks from secondary particles from the
fit region to the region $|\dzerobs| < 1.5$~mm. 
Different generators are used to estimate the interpolation and differences between data and simulation in the shape of the $\dzerobs$ distribution in the fit region are considered.
Additional, much smaller, systematic uncertainties arise from a variation of the fit range,
considering the $\eta$ dependence of the fitted fractions and from using
special simulation samples with varying amounts of detector material.

There is a second source of non-primary particles:  charged particles with  a
 mean lifetime $30 < \tau < 300$~ps which, unlike in previous analyses~\cite{STDM-2010-06}, are excluded from the primary-particle definition. 
These are charged strange baryons that decay after a short
flight length and have a very low track reconstruction efficiency.
Reconstructed tracks from these particles
are treated as background and are subtracted.
The fraction of
reconstructed tracks coming from strange baryons is estimated from
simulation with \epos{}
 to be $(0.01 \pm 0.01)\%$ on average, with the fraction increasing with track
\pta\ to be $(3 \pm 1)\%$ above 20~GeV. 
The fraction is much smaller at low \pta{} due to the extremely low efficiency
of reconstructing a track from a particle that decays early in the
detector. 
 The systematic uncertainty
is taken as the maximum difference between the nominal \epos{} prediction
and that of \py{} \atwo{} or \py{} \monash{}, which is then symmetrised.

\section{Selection efficiency}\label{sec:eff-corr}

The data are corrected to obtain inclusive spectra for primary charged
particles satisfying the particle-level kinematic requirements. 
These corrections account for inefficiencies due to trigger selection, 
vertex and track reconstruction. 

In the following sections the methods used to obtain these efficiencies, as well as the systematic uncertainties associated with them, are described. 

\subsection{Trigger efficiency}\label{sec:trigeff}

\begin{figure*}
	\begin{center}
	\subfigure[]{
		\includegraphics[width=0.48\textwidth]{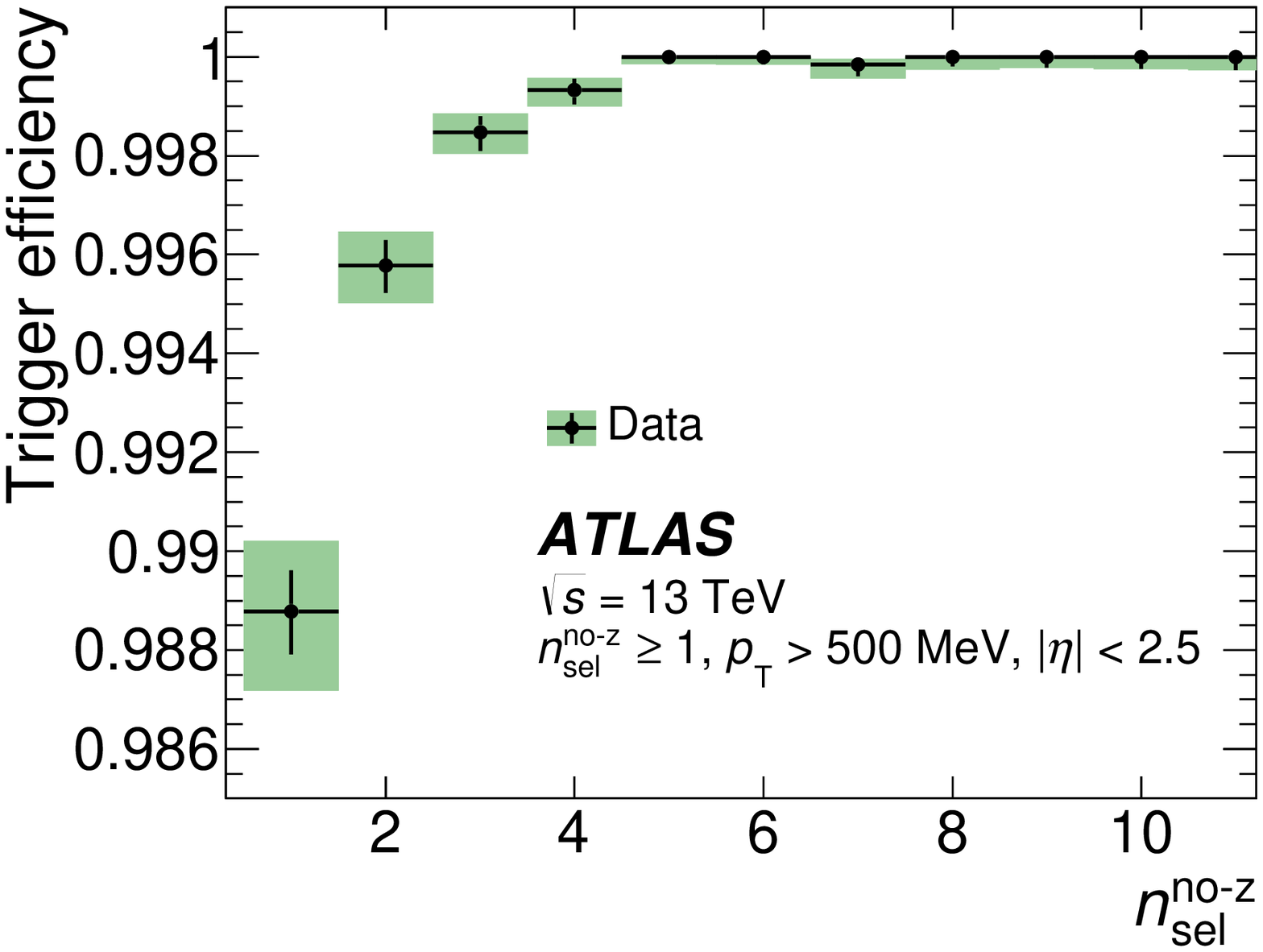}
                \label{fig:eff-trigger}
	}
	\subfigure[]{

		\includegraphics[width=0.48\textwidth]{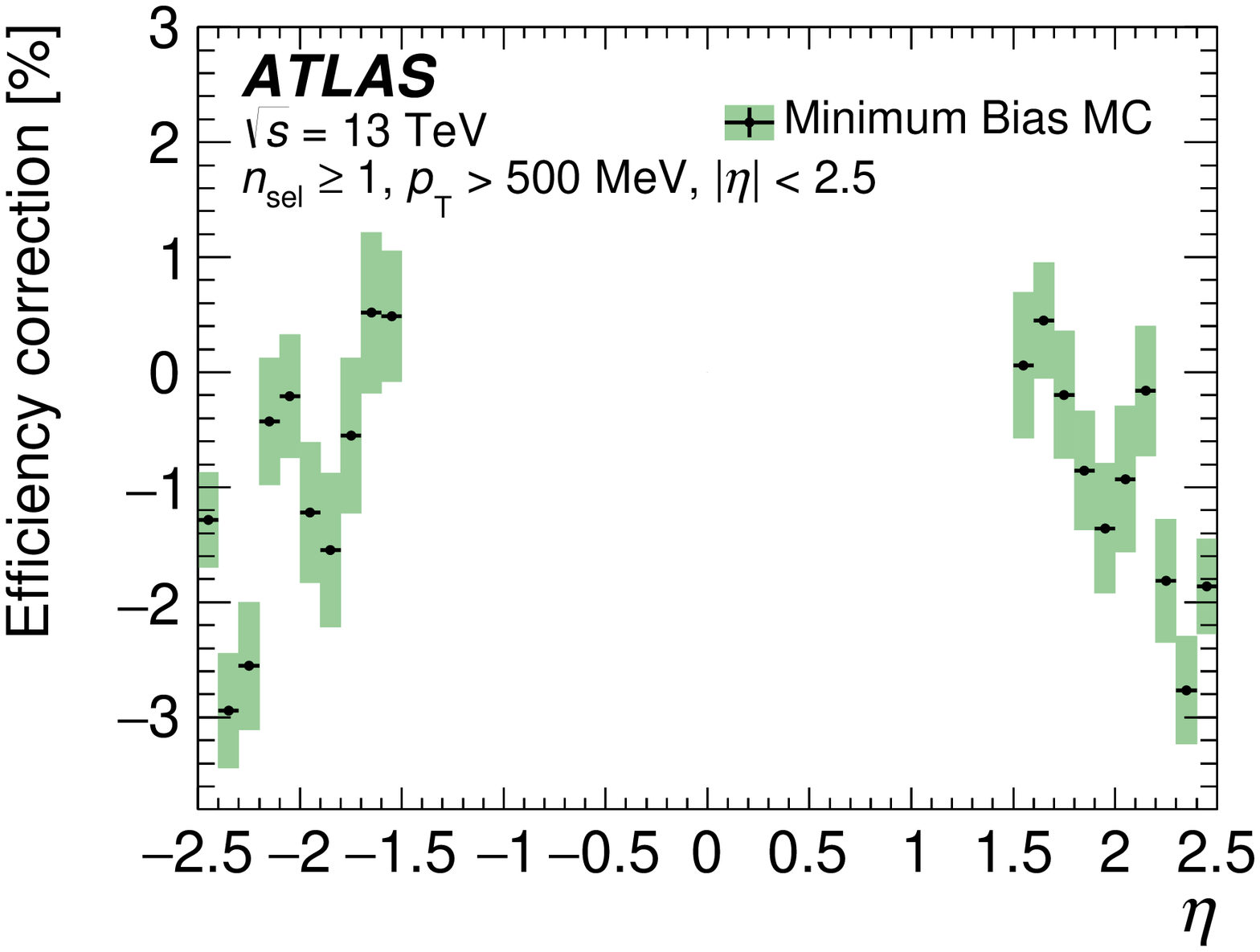}
                \label{fig:eff-corr}
	}
	\subfigure[]{
		\includegraphics[width=0.48\textwidth]{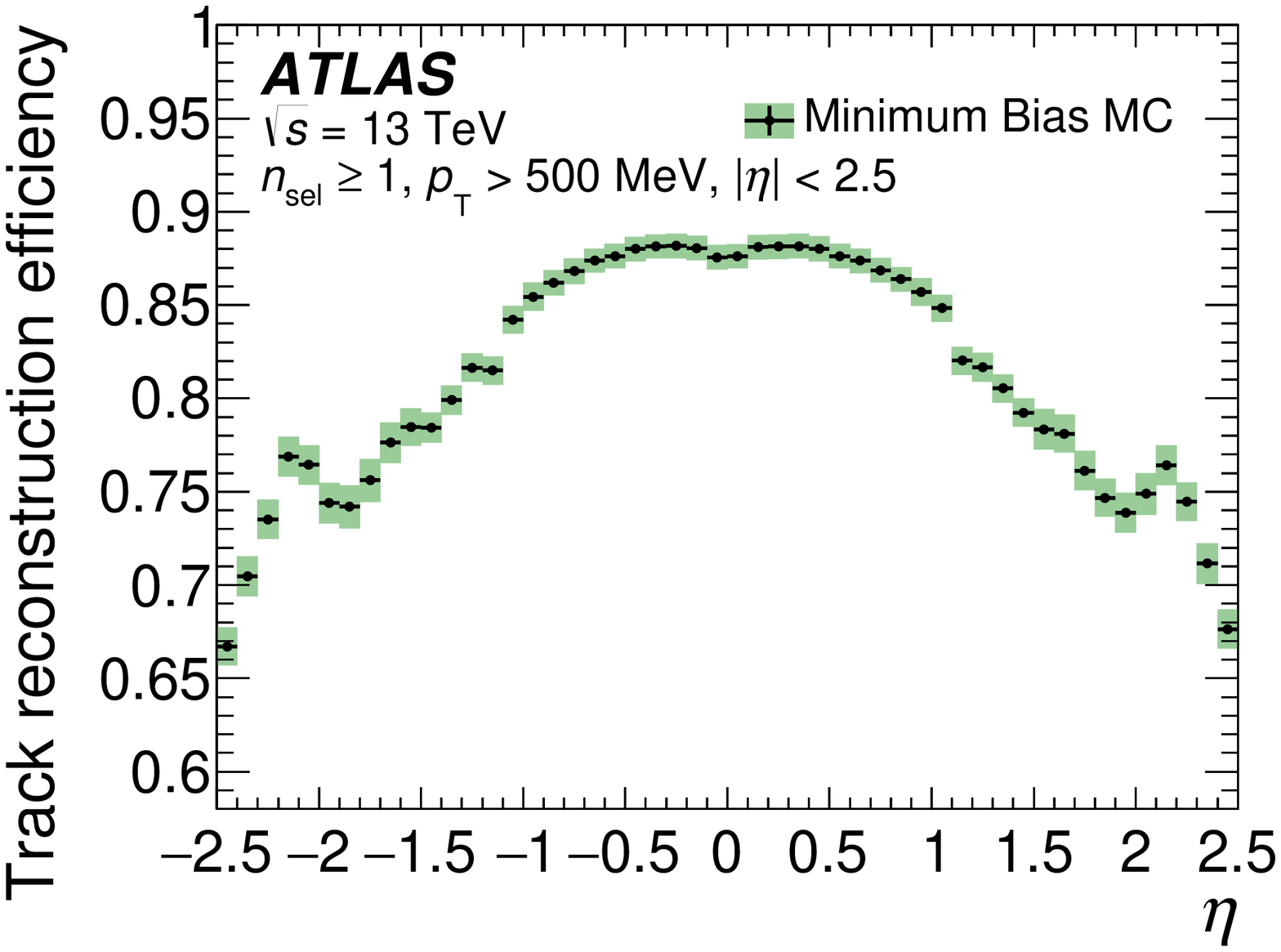}
                \label{fig:eff-tracketa}
	}
	\subfigure[]{
		\includegraphics[width=0.48\textwidth]{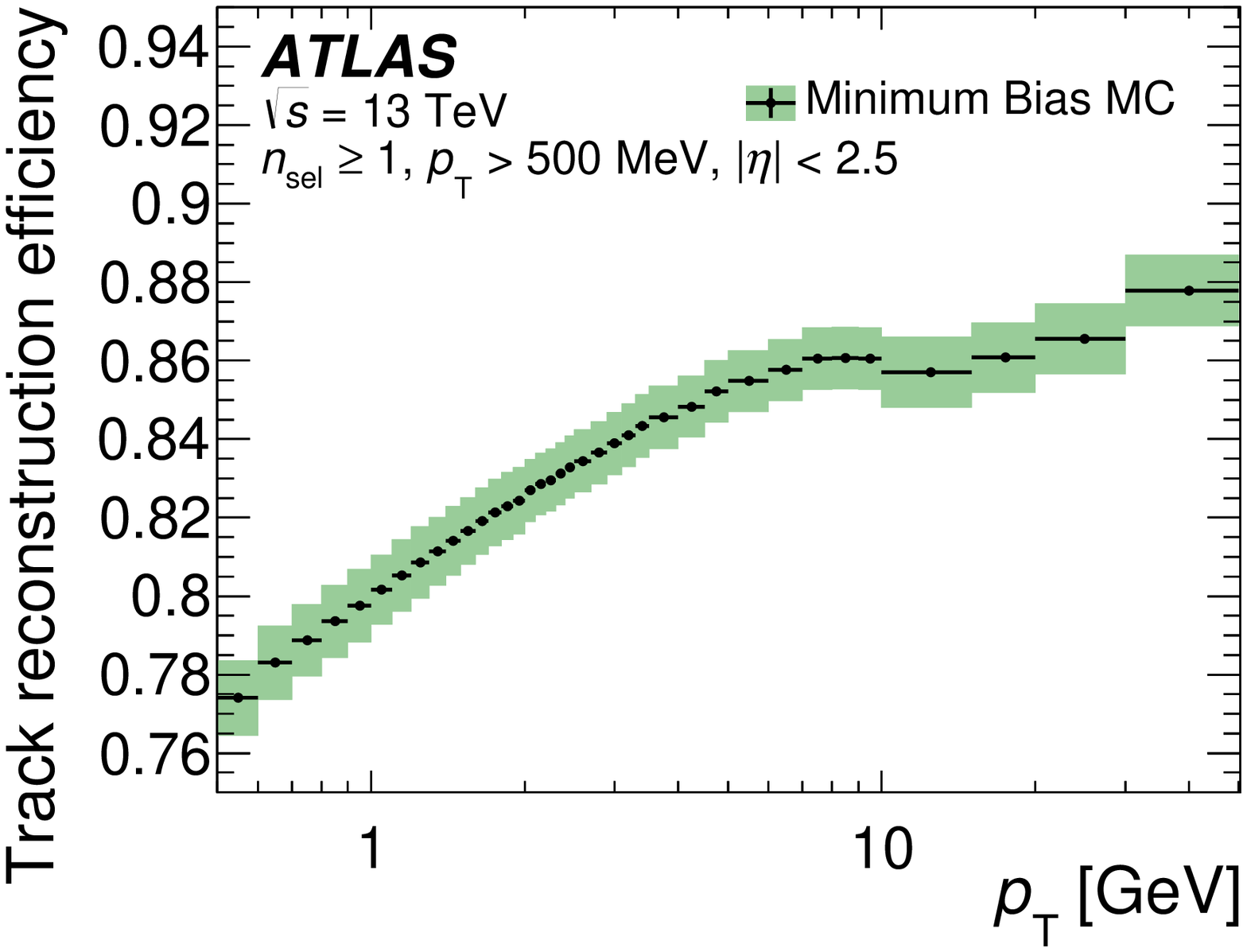}
                \label{fig:eff-trackpt}
	}
	\end{center}
	\caption {(a) Trigger efficiency with respect to the event selection, 
as a function of the number of reconstructed tracks without the
$\zzsint$  constraint (\nselbs).
(b) Data-driven correction to the track  reconstruction efficiency as
a  function of pseudorapidity, $\eta$.
The track reconstruction efficiency after this correction as a function of (c)
$\eta$ and
(d) transverse momentum, \pta\ as predicted by \py\ \atwo\ and single-particle simulation.
The statistical uncertainties are shown as black vertical bars, the total uncertainties  as green shaded areas.}
  \label{fig:eff}
\end{figure*}

The trigger efficiency, \trigeff , is measured from a data sample
selected using the control trigger described in Section~\ref{sec:detector}.
The requirement of an event primary vertex is removed for these
trigger studies, to account for possible correlations between the trigger and
vertex reconstruction efficiencies.
The trigger efficiency is therefore parameterised as a function of \nselbs, which
is defined as the number of tracks passing all of the track
selection requirements except for the $\zzsint$  constraint, as this
requires knowledge of the primary vertex position.
The trigger efficiency is taken to be the fraction of events from the control trigger 
in which the MBTS trigger also accepted the event.
This is shown in Figure~\ref{fig:eff-trigger} as a function of
\nselbs. 
The efficiency is measured to be just below 99\% for $\nselbs=1$ and it rapidly rises to 100\% at higher track multiplicities.
The trigger requirement is found to introduce no observable bias in
the \pta\ and $\eta$ distributions of selected tracks.
Systematic uncertainties are estimated
from differences in the 
trigger efficiency measured on each of the two sides of the
detector 
and
 from a study that assesses the impact of beam-induced background and
 tracks from secondary particles by varying the impact parameter requirements on selected
 tracks. The total systematic uncertainty is $\pm 0.15\%$\ for $\nselbs=1$ and
 it rapidly decreases at higher track multiplicities. This uncertainty
 is negligible compared to those from other sources and is therefore neglected.
\subsection{Vertex reconstruction efficiency}
 \label{sec:vertexeff}

The vertex reconstruction efficiency, \vtxeff, is determined from data
by taking the ratio of the number of selected events with a reconstructed vertex to
the total number of events with the requirement of a primary vertex
removed.
The expected
contribution from beam background events is estimated using the same
method as described in Section~\ref{sec:background} and subtracted before
measuring the efficiency.
Like the trigger efficiency, the vertex efficiency is measured in bins
of \nselbs\ as the \zzsint\ constraint cannot be applied to the tracks
in this study.
The efficiency is measured to be just below 90\% for $\nselbs=1$ and
it rapidly rises to 100\% at higher track multiplicities.
In events with $\nselbs=1$ the efficiency is also measured as a
function of $\eta$ of the track, and the efficiency increases monotonically from 81\% at $|\eta|  = 2.5$ to 93\% at  $|\eta|  = 0$.
The systematic uncertainty is estimated from 
the difference between the vertex reconstruction efficiency measured
prior to and after beam background removal.
The uncertainty is $\pm 0.1\%$\ for $\nselbs=1$ and
rapidly decreases at higher track multiplicities.
This uncertainty
 is negligible compared to those from other sources and is therefore neglected.
\subsection{Track reconstruction efficiency}\label{sec:trkeff}

The primary track reconstruction efficiency, \trkeff , is determined from
the simulation, corrected to account for differences
between data and simulation in
the amount of detector material between the pixel and SCT detectors in
the region $|\eta| > 1.5$. In the other regions of the detector there is an uncertainty due to the knowledge of the detector material that will be discussed below, but no correction is applied.
The efficiency is parameterised in two-dimensional bins of \pta\ and
$\eta$ and is defined as:
$$
\trkeff (p_\mathrm{T},\eta) \ = \ \frac{N^\mathrm{matched}_\mathrm{rec}(p_\mathrm{T},\eta)}
                                   {N_\mathrm{gen}(p_\mathrm{T},\eta)},
$$
where  \pta\ and $\eta$ are generated particle properties,
 $N^\mathrm{matched}_\mathrm{rec}(p_\mathrm{T},\eta)$ is the number of
 reconstructed tracks matched to a generated primary charged particle  
and $N_\mathrm{gen}(p_\mathrm{T},\eta)$ is the number of generated
primary charged particles in that bin. 
A track is matched to a generated particle if the weighted fraction of hits on
the track which originate from that particle exceeds 50\%.
The hits are weighted such that all subdetectors have the same weight
in the sum.

The track reconstruction efficiency depends on the amount of material in the
detector, due to particle interactions that lead to efficiency losses.
The relatively large amount of material between the pixel and SCT detectors in the
region $|\eta|>1.5$ has changed between Run~1 and Run~2 due to the
replacement of some pixel services, which are
difficult to simulate accurately.
 The
track reconstruction efficiency in this region is corrected using a method
that compares the efficiency to extend a track reconstructed
  in the pixel detector into the SCT in data and
  simulation. Differences in this extension efficiency are sensitive
  to differences in the amount of material in this region. The
  correction together with the systematic uncertainty,  coming predominantly from the uncertainty of the
particle composition in the simulation used to make the measurement, is shown in Figure~\ref{fig:eff-corr}.
The uncertainty in the track reconstruction efficiency resulting from this
correction is $\pm  0.4\%$\ in the region  $|\eta|>1.5$.

The resulting reconstruction efficiency as a function of $\eta$ integrated over \pta\ is
shown in Figure~\ref{fig:eff-tracketa}.
The track reconstruction efficiency is lower in the region $|\eta| >
1$ due to particles passing through more material in that region.
The slight increase in efficiency at $|\eta|\sim2.2$ is due to the
particles passing through an increasing number of layers in the ID end-cap.
Figure~\ref{fig:eff-trackpt} shows the efficiency as a function of
\pta\ integrated over $\eta$.


A good description of the material in the detector in the regions not
probed by the method described above (which only probes the material between the pixel and SCT detectors in
the region $|\eta| > 1.5$) 
is needed to
obtain a good description of the track reconstruction efficiency.
The material within the ID was studied extensively during Run
1~\cite{PERF-2011-08},  where the amount of material was known to within
$\pm 5\%$. This gives rise to a systematic uncertainty in the track
reconstruction efficiency of $\pm0.6\% \, (\pm1.2\%)$ in the
most central (forward) region.
Between Run 1 and Run 2 the IBL was installed, the simulation of
which must therefore be studied with the Run 2 data.
Two data-driven methods are used: a study of secondary vertices from photon conversions
  ($\gamma\rightarrow e^{+}e^{-}$) and a study of secondary vertices from 
  hadronic interactions, where the radial position of the vertex is measured with good precision.
Comparisons between data and simulation indicate that the material in
the IBL is constrained to within $\pm 10\%$. This leads to an uncertainty in
the track reconstruction efficiency of $\pm 0.1\%$\ ($\pm 0.2\%$) in the
central (forward) region. This uncertainty is added linearly
to the uncertainty from constraints from Run 1, to cover the
possibility of missing material in the simulation
in both cases. The resulting 
uncertainty is added in quadrature to the uncertainty from the
data-driven correction. 
The total uncertainty due to the imperfect knowledge of the detector
material is $\pm 0.7\%$\ in the
most central region and $\pm 1.5\%$\ in the most forward region.

There is a small difference in efficiency,  between data and simulation,  of the requirement that each 
reconstructed track has at least one pixel hit, at least
six SCT hits, an innermost-pixel-layer hit if expected 
(if a hit in the innermost layer is not expected, the next-to-innermost hit
is required if expected) and a track-fit $\chi^2$ probability  greater than $0.01$ for tracks with $\pta~>~10$~GeV.  
This difference is assigned as a further systematic
uncertainty, amounting to $\pm 0.5\%${} for $\pta < 10$~GeV  and $\pm 0.7\%${}
for $\pta > 10$~GeV. 

The total uncertainty due to the track reconstruction efficiency
determination, shown in Figures~\ref{fig:eff-tracketa}
and~\ref{fig:eff-trackpt}, is obtained by adding all effects in
quadrature and is dominated by the uncertainty from the material description.

\section{Correction procedure}\label{sec:corr}
The following steps are taken to correct the measurements for detector effects.

\begin{itemize}
\item All distributions are corrected for the loss of events due to the trigger and
vertex requirements by reweighting events
according to the function:

$$
w_\mathrm{ev}(\nselbs, \eta) =  \frac{1}{\trigeff (\nselbs)} \cdot
\frac{1}{\vtxeff (\nselbs, \eta)},
$$
where the $\eta$ dependence is only relevant for $\nselbs=1$,
as discussed in Section~\ref{sec:vertexeff}.

\item The  $\eta$  and \pta\ distributions of selected tracks are corrected using a track-by-track weight:

$$
w_\mathrm{trk}(\pta, \eta) =  \frac{1-\fsec( \pta, \eta ) - \fsb( \pta ) -\fokr(\pta , \eta)}{\trkeff( \pta,\eta )}
$$
where \fsec\ and \fsb\  are the fraction of tracks from secondary particles and
 from strange baryons respectively, determined as described in Section~\ref{sec:background}. 
The fraction of selected tracks for
which the corresponding primary particle is outside the kinematic
range, $\fokr(\pta, \eta)$, originates from resolution effects and is 
estimated from the simulation  to be 3.5\%{}  at $\pta=500$~MeV, decreasing to 1\%{} for $\pta=1$~GeV and is only relevant for $2.4 < |\eta| < 2.5$.
No additional corrections are needed for the $\eta$  distribution.
For the \pta\ distribution a Bayesian
unfolding~\cite{D'Agostini:1994zf}  is applied to correct the measured
track \pta\ distribution to that for primary particles.

\item After applying the trigger and vertex efficiency corrections, the
Bayesian unfolding is applied to the multiplicity distribution in order
to correct from the observed track multiplicity to the multiplicity of
primary charged particles, and therefore the track reconstruction efficiency weight does not need to be applied. The correction procedure also accounts for events
that have migrated out of the selected kinematic range  ($\nch\geq1$).

\item The total number of events, \nev, used to normalise the distributions, is defined as the integral of the \nch\ distribution, after all corrections are applied.

\item 
The dependence of \meanpt{} on \nch{} is obtained by first separately correcting $\sum_i \pta(i)$ (summing over the \pta\ of all tracks and all events) versus the number of selected tracks and the total number of tracks in all events versus the number of selected tracks, and then taking the ratio.
They are corrected using the appropriate track
weights first, followed by the Bayesian unfolding procedure.
\end{itemize}

Systematic uncertainties in the track reconstruction efficiency, discussed in
Section~\ref{sec:eff-corr}, and the
fraction of tracks from non-primary particles, discussed in
Section~\ref{sec:background}, give rise to an uncertainty in
$w_\mathrm{trk}(\pta, \eta)$, directly affecting 
 the $\eta$ and \pta{} distributions. 
 For the \nch\ distribution, where the track weights are not explicitly applied, 
the effects from uncertainties in these sources are found by modifying the 
distribution of selected tracks in data.  In each multiplicity interval tracks are 
randomly removed or added with probabilities dependent on the uncertainties
in the track weights of tracks populating that bin. 
This modified distribution is then unfolded
and the deviation from the nominal \nch\ distribution is
taken as a systematic uncertainty. 
An  uncertainty from the fact that the correction
procedure, when applied to simulated events, does not reproduce exactly
the distribution from generated particles (non-closure) is included in
all measurements.
An additional systematic uncertainty in the measured \pta~distribution
arises from possible biases and degradation in the
 \pta\ measurement. This is quantified by comparing the track hit
 residuals in data and simulation.
The effectiveness of the track-fit $\chi^2$ probability selection in
suppressing tracks reconstructed with high momentum but originating from
low momentum particles was also considered; it was found that the
fraction of these tracks remaining was consistent with predictions
from simulation. An uncertainty due to the statistical precision of
the check is included for the \pta~distribution. 
Uncertainty sources that also affect \nev{} partially cancel in
the final distributions.
A summary of the main systematic uncertainties affecting the $\eta$,
\pta{} and \nch{} distributions  is given in Table~\ref{tab:systematics}.
\begin{table}
\begin{center}
	\caption{Summary of systematic uncertainties on the $\eta$,
\pta{} and \nch{} distributions.}
	\begin{tabular}{lcc} 
          \hline 
          Source            &  Distribution
          & Range of values \\
          \hline  
          
           Track reconstruction efficiency                                  & $\eta$                  &  0.5\%{}
                               --   1.4\% \\
                                            & \pta{}
                                            & 0.7\%    \\
     & \nch{}                   &   0\%{}
        -- $^{+17\%}_{-14\%}$  \\
         
      Non-primaries                         & $\eta$                  & 0.5\%   \\
                               & \pta{}                    & 0.5\%  --
                               0.9\% \\       
  & \nch{}                    &  0\%{} -- $^{+10\%}_{\,\,\,-8\%}$ \\
       
        Non-closure       & $\eta$                  & 0.7\%   \\
                               & \pta{}                    & 0\%{} -- 2\%
                               \\   
& \nch{}                   &  0\%{} -- 4\%{}  \\
          \pta{}-bias     & \pta{}                    & 0\%{} --
          5\%    \\   
 High-\pta{}     & \pta{}                    & 0\%{} -- 1\%    \\          
         \hline
	\end{tabular}
        \label{tab:systematics}
\end{center}
\end{table}

Uncertainties in the \meanptN\
measurement are found in the same way as those in the \nch\ distribution.
The dominant uncertainty is from non-closure
which varies from $\pm 2\%$\ at low \nch\
to $\pm 0.5\%$\ at high \nch.
All other uncertainites largely cancel in the ratio and are negligible.
At high \nch{} the total uncertainty is dominated by the statistical
uncertainty.

\section{Results}
\label{sec:result}
\begin{figure*}
	\begin{center}
	\subfigure[]{
		\includegraphics[width=0.43\textwidth]{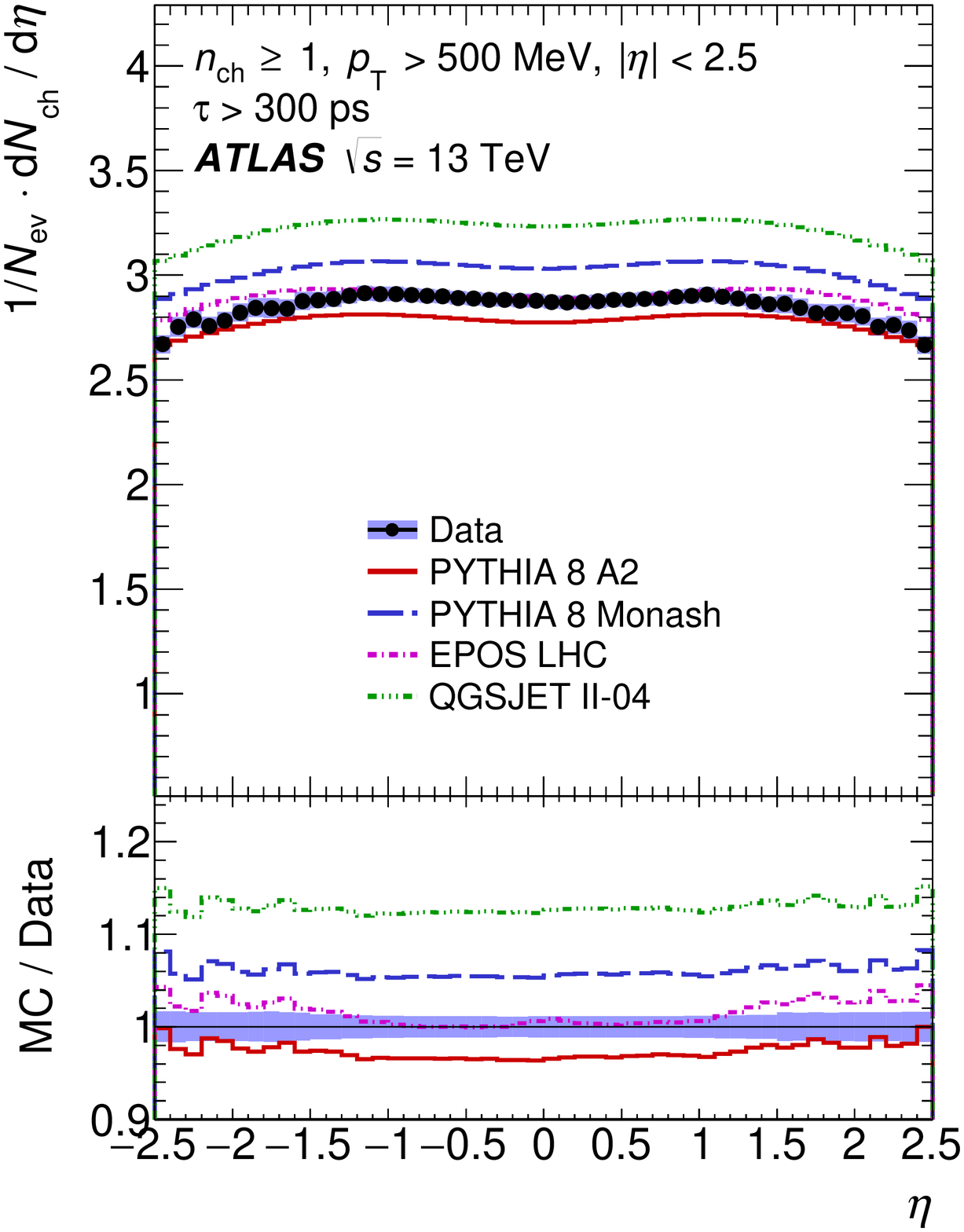}
                \label{fig:dndeta}
	}
	\subfigure[]{
		\includegraphics[width=0.43\textwidth]{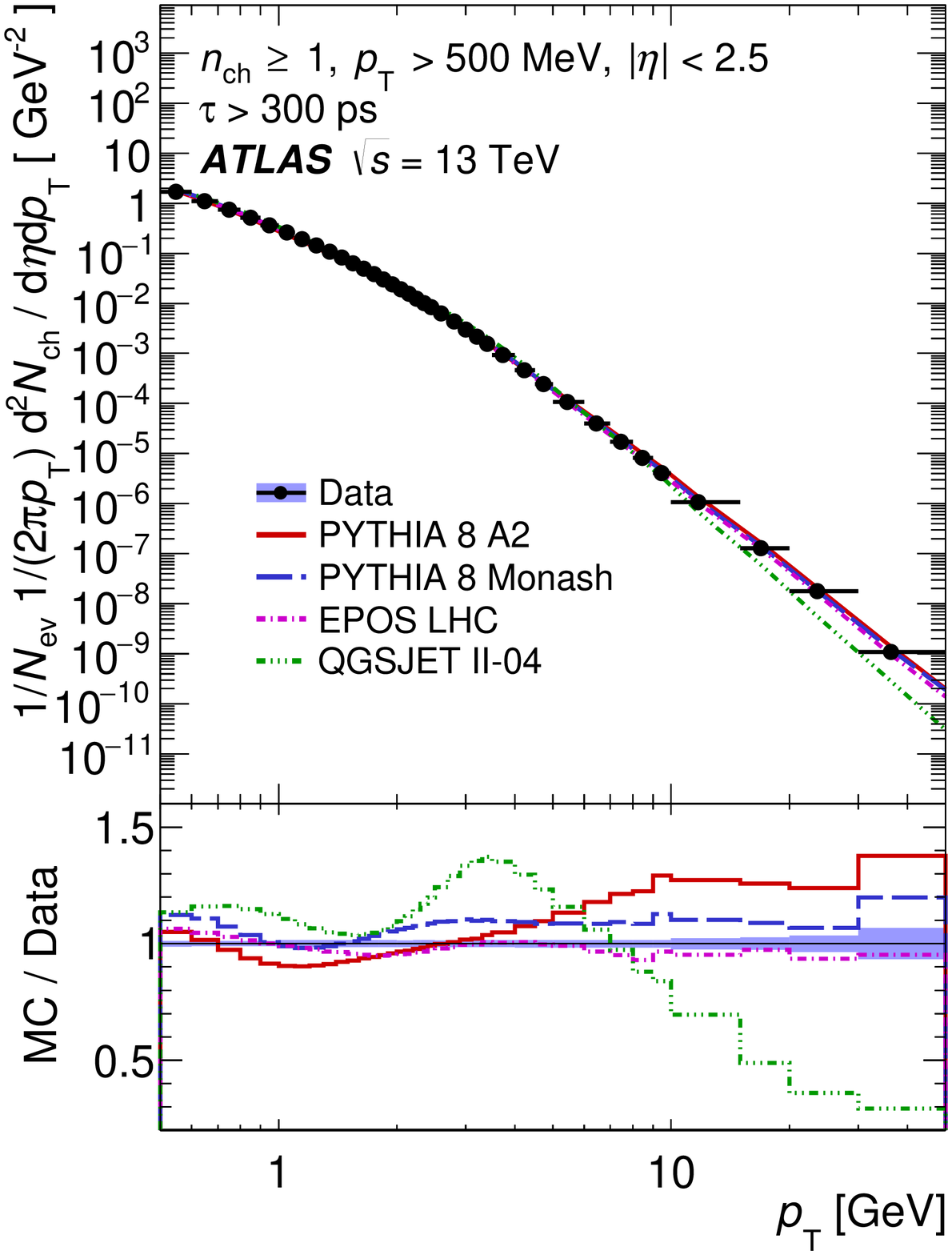}
                \label{fig:dndpt}
	}
	\subfigure[]{
		\includegraphics[width=0.43\textwidth]{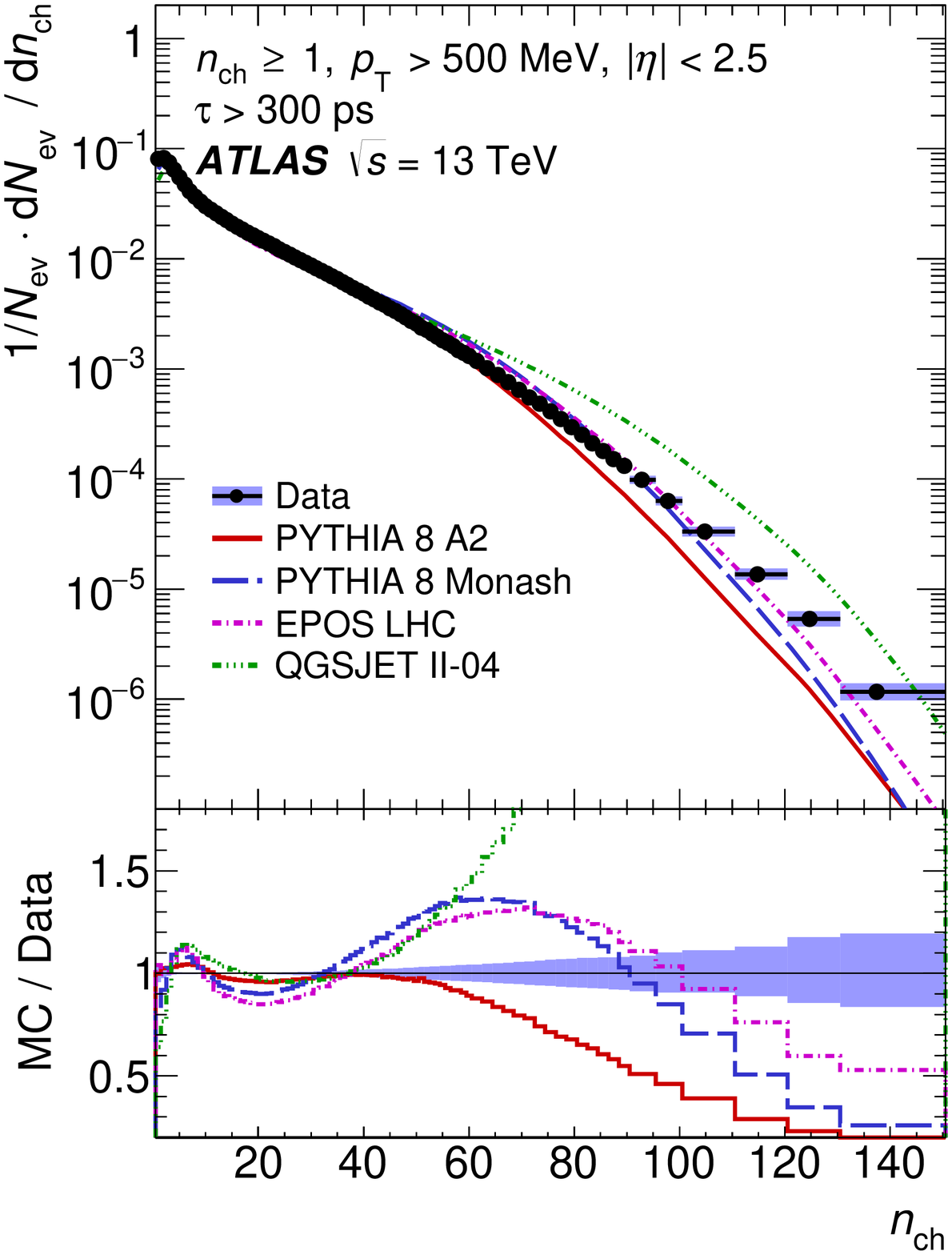}
                \label{fig:dNdnch}
	}
	\subfigure[]{
		\includegraphics[width=0.43\textwidth]{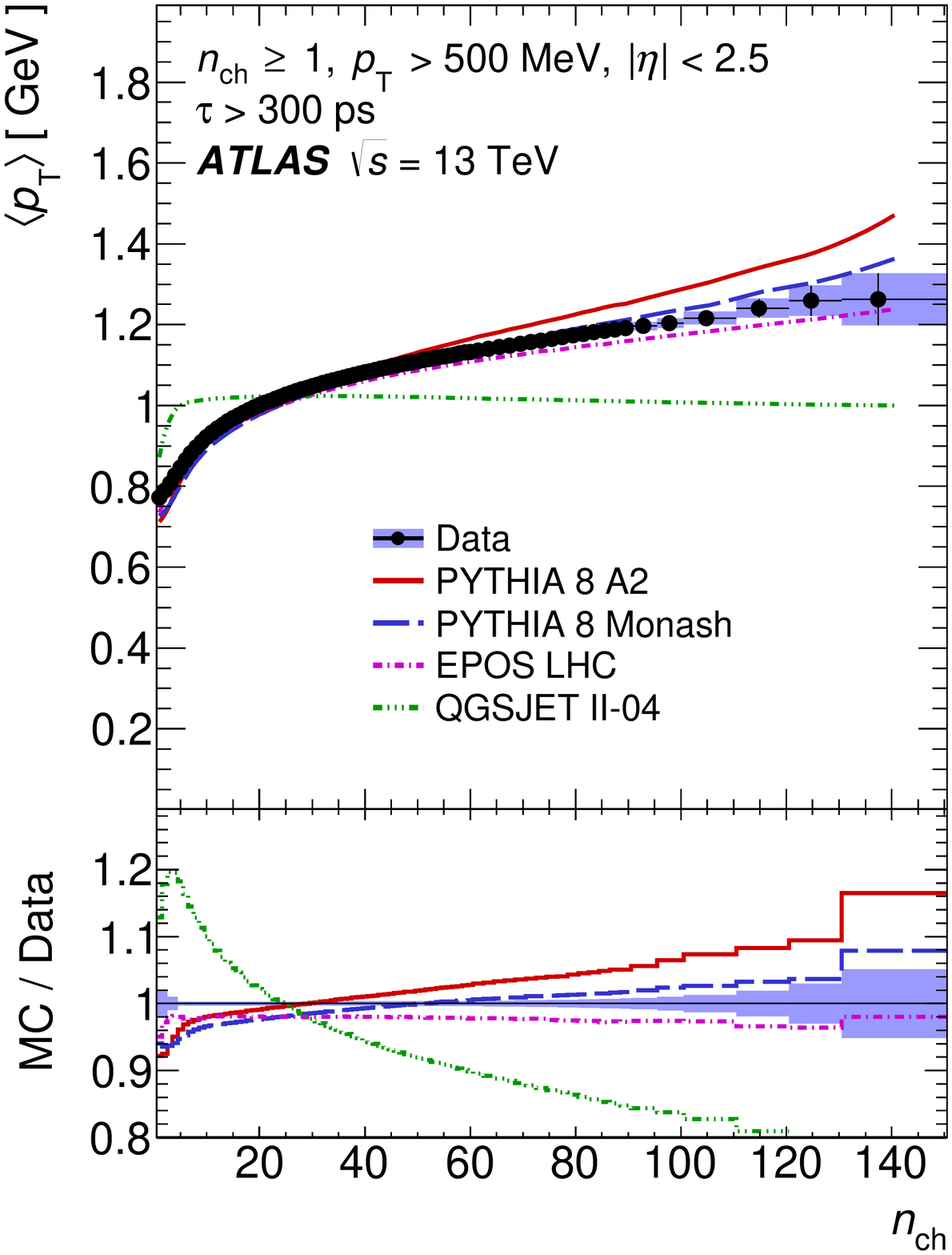}
                \label{fig:meanpt}
	}
	\end{center}
	\caption {Primary-charged-particle multiplicities as a function of 
          (a) pseudorapidity, $\eta$, and (b) transverse momentum, \pta{}; (c) the
          multiplicity, \nch{}, distribution and (d) the mean transverse
          momentum, \meanpt{} , versus \nch{}
  in events with $\nch \geq 1$, $\pta > 500$~MeV and $|\eta| < 2.5$. 
The dots represent the data and the curves the predictions from different MC
models. The $x$-value in each bin corresponds to the bin centroid. The
vertical bars represent the statistical uncertainties, while the shaded
areas show statistical and systematic uncertainties added in quadrature. The
bottom panel in each figure shows the ratio of the MC simulation to data. Since the bin centroid is different for data and simulation, the values of the
ratio correspond to the averages of the bin content.
}
  \label{fig:results}
\end{figure*}

The corrected distributions for primary charged particles in events
with $\nch \geq 1$ in the kinematic range $\pta > 500$~MeV and $|\eta|
< 2.5$ are shown in Figure~\ref{fig:results}. 
In most regions of all distributions the dominant uncertainty comes from the track reconstruction efficiency.
The results are compared to predictions of models tuned to a wide
range of measurements. 
The measured distributions are presented as inclusive distributions with corrections that rely minimally on the MC model used, in order to facilitate an accurate comparison with predictions.

Figure~\ref{fig:dndeta} shows the multiplicity of charged particles as a function of pseudorapidity. 
The mean particle density is roughly constant at 2.9 for $|\eta|<1.0$ and
decreases at higher values of $|\eta|$. \epos\ describes the data for
$|\eta|<1.0$, and predicts a slightly larger multiplicity at larger $|\eta|$ values. \qgsjet\ and \py\ \monash\
predict multiplicities that are too large by  approximately 15\%\ and 5\%\ respectively. 
\py\ \atwo\ predicts a multiplicity that is 3\%\ too low in the central
region, but describes the data well in the forward region.

Figure~\ref{fig:dndpt} shows the charged-particle transverse momentum distribution. \epos\ 
describes the data well over the entire \pta\ spectrum. The \py\ tunes describe the data
reasonably well, but are slightly  above the data in the high-\pta\ region. 
\qgsjet\ gives a poor prediction over the entire spectrum,
overshooting the data
in the low-\pta\ region and undershooting it in the
high-\pta\ region. 

Figure~\ref{fig:dNdnch} shows the charged-particle multiplicity
distribution.  The high-\nch{} region has significant contributions
from events with numerous MPI.
\py\ \atwo\ describes the data in the region $\nch
< 50$, but predicts too few events at larger $\nch$ values. 
\py\ \monash, \epos\ and \qgsjet\ describe the data reasonably well in
the region $\nch < 30$ but predict too many events in the mid-$\nch$
region, with \py\ \monash\ and \epos\ predicting too few events in the
region $\nch > 100$ while \qgsjet\ continues to be above the data.

Figure~\ref{fig:meanpt} shows the mean transverse momentum versus the
 charged-particle multiplicity.
The \meanpt\ rises with \nch, from $0.8$ to $1.2$~GeV. This increase
is expected due to colour coherence effects being important in dense
parton environments and is modelled by a colour reconnection mechanism in
\py\ or by the hydrodynamical evolution model used in \epos. 
If the high-\nch\ region is assumed to be dominated by events with
 numerous MPI, without colour coherence effects the \meanpt\
is approximately independent of \nch. Including colour coherence effects
leads to fewer additional charged particles produced with every
additional MPI, with an equally
large \pta\ to be shared among the produced
hadrons~\cite{Sjostrand:2013cya}.
\epos\ predicts a slightly lower \meanpt, but describes the
dependence on \nch\ very well. The \py\ tunes predict a steeper
rise of \meanpt\ with $\nch$ than the data, predicting lower values in the
low-\nch\ region and higher values in the high-\nch\ region.
\qgsjet\ predicts a \meanpt\ of $\sim1$~GeV,
with very little dependence on \nch; this is expected as it contains
no model for colour coherence effects.

In summary, \epos\ and the \py\ tunes describe the data most
accurately, with \epos\ reproducing the $\eta{}$ and \pta{} distributions
and the
\meanptN\ the best and \py\ \atwo\ describing the
multiplicity the best in the low- and mid-$\nch$ regions. \qgsjet\
provides an inferior description of the data.


\begin{figure}[htb!]
\begin{center}
\includegraphics[width=0.48\textwidth]{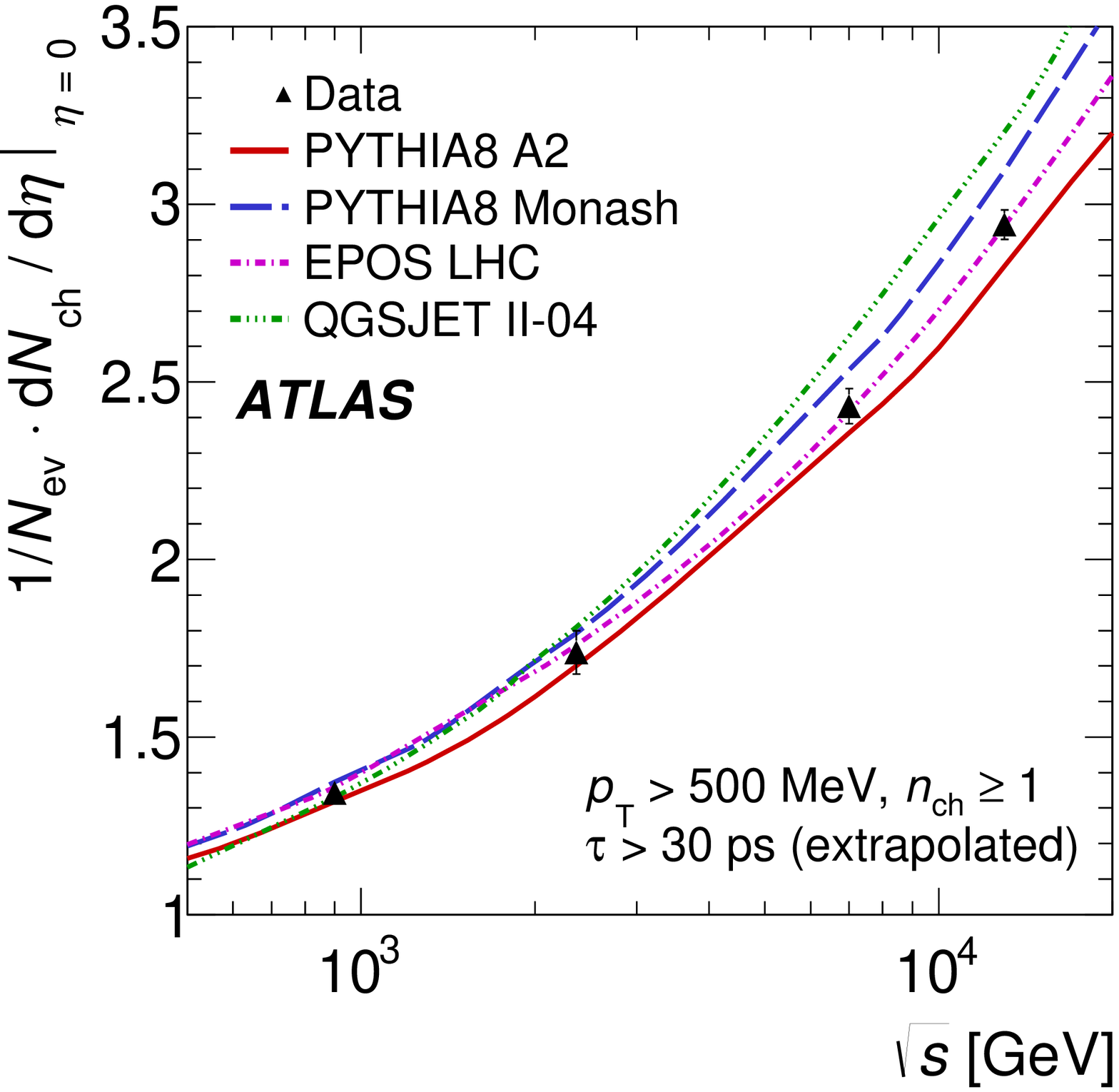}
\caption{The average primary-charged-particle multiplicity in $pp$ interactions per unit of
  pseudorapidity, $\eta$, for $|\eta|~<~0.2$ as a function of the centre-of-mass energy. The values at centre-of-mass energies other than 13~\TeV are taken from Ref.~\cite{STDM-2010-06}. 
Charged strange baryons are included in the definition of primary particles. 
The data are compared to various particle-level MC predictions. 
The vertical error bars on the data represent the total uncertainty.}
\label{fig:sqrts}
\end{center}
\end{figure}

The mean number of primary charged particles in the central region is computed
by averaging over $|\eta| < 0.2$ to be 2.874 $\pm$ 0.001\stat{} $\pm$
0.033\syst{}. This measurement is then corrected for the contribution from strange baryons
 and compared to previous
measurements~\cite{STDM-2010-06} at different $\sqrt{s}$ values in
Figure~\ref{fig:sqrts} together with the MC predictions. 
The correction factor for strange baryons depends on the MC model used and is found to be $1.0241 \pm 0.0003$ (\epos),
$1.0150 \pm 0.0004$ (\py\ \monash) and $1.0151 \pm 0.0002$ (\py\ \atwo),
where the uncertainties are statistical.
\qgsjet\ does not include charged strange baryons.
The prediction from \epos\ is used to
perform the extrapolation
and the deviation from the \py\ \monash{} prediction is taken as a systematic
uncertainty and symmetrised to give $1.024 \pm 0.009$.

The mean number of primary charged particles increases by a factor of 2.2 when
$\sqrt{s}$ increases by a factor of about 14 from 0.9~TeV to 13~TeV.
\epos\ and \py\ \atwo\ describe the dependence on $\sqrt{s}$ very
well, while \py\ \monash\ and \qgsjet\ predict a steeper rise in
multiplicity with $\sqrt{s}$.


\section{Conclusion}
\label{sec:conclusion}
Primary-charged-particle multiplicity measurements with the ATLAS detector
using proton--proton collisions delivered by the LHC at
$\sqrt{s}=13$~TeV are presented.
From a data sample corresponding to an integrated luminosity of 170~$\mu$b$^{-1}$, nearly nine million inelastic interactions with at least one
reconstructed track with $|\eta| < 2.5$ and $\pta > 500$~MeV are analysed.
The results highlight clear differences between MC models and the measured distributions.  
Among the models considered \epos\ reproduces the data the best, \py\
\atwo\ and \monash\
give reasonable descriptions of the data and
\qgsjet\ provides the worst description of the data.

\subsection*{Acknowledgements}

We thank CERN for the very successful operation of the LHC, as well as the
support staff from our institutions without whom ATLAS could not be
operated efficiently.

We acknowledge the support of ANPCyT, Argentina; YerPhI, Armenia; ARC, Australia; BMWFW and FWF, Austria; ANAS, Azerbaijan; SSTC, Belarus; CNPq and FAPESP, Brazil; NSERC, NRC and CFI, Canada; CERN; CONICYT, Chile; CAS, MOST and NSFC, China; COLCIENCIAS, Colombia; MSMT CR, MPO CR and VSC CR, Czech Republic; DNRF, DNSRC and Lundbeck Foundation, Denmark; IN2P3-CNRS, CEA-DSM/IRFU, France; GNSF, Georgia; BMBF, HGF, and MPG, Germany; GSRT, Greece; RGC, Hong Kong SAR, China; ISF, I-CORE and Benoziyo Center, Israel; INFN, Italy; MEXT and JSPS, Japan; CNRST, Morocco; FOM and NWO, Netherlands; RCN, Norway; MNiSW and NCN, Poland; FCT, Portugal; MNE/IFA, Romania; MES of Russia and NRC KI, Russian Federation; JINR; MESTD, Serbia; MSSR, Slovakia; ARRS and MIZ\v{S}, Slovenia; DST/NRF, South Africa; MINECO, Spain; SRC and Wallenberg Foundation, Sweden; SERI, SNSF and Cantons of Bern and Geneva, Switzerland; MOST, Taiwan; TAEK, Turkey; STFC, United Kingdom; DOE and NSF, United States of America. In addition, individual groups and members have received support from BCKDF, the Canada Council, CANARIE, CRC, Compute Canada, FQRNT, and the Ontario Innovation Trust, Canada; EPLANET, ERC, FP7, Horizon 2020 and Marie Sk{\l}odowska-Curie Actions, European Union; Investissements d'Avenir Labex and Idex, ANR, Region Auvergne and Fondation Partager le Savoir, France; DFG and AvH Foundation, Germany; Herakleitos, Thales and Aristeia programmes co-financed by EU-ESF and the Greek NSRF; BSF, GIF and Minerva, Israel; BRF, Norway; the Royal Society and Leverhulme Trust, United Kingdom.

The crucial computing support from all WLCG partners is acknowledged
gratefully, in particular from CERN and the ATLAS Tier-1 facilities at
TRIUMF (Canada), NDGF (Denmark, Norway, Sweden), CC-IN2P3 (France),
KIT/GridKA (Germany), INFN-CNAF (Italy), NL-T1 (Netherlands), PIC (Spain),
ASGC (Taiwan), RAL (UK) and BNL (USA) and in the Tier-2 facilities
worldwide.

\clearpage

\printbibliography

\phantom{x}
\newpage

\appendix
\part*{Appendix}
\section{Results in a common phase space}
\label{sec:reduced}
\begin{figure*}
	\begin{center}
	\subfigure[]{
		\includegraphics[width=0.43\textwidth]{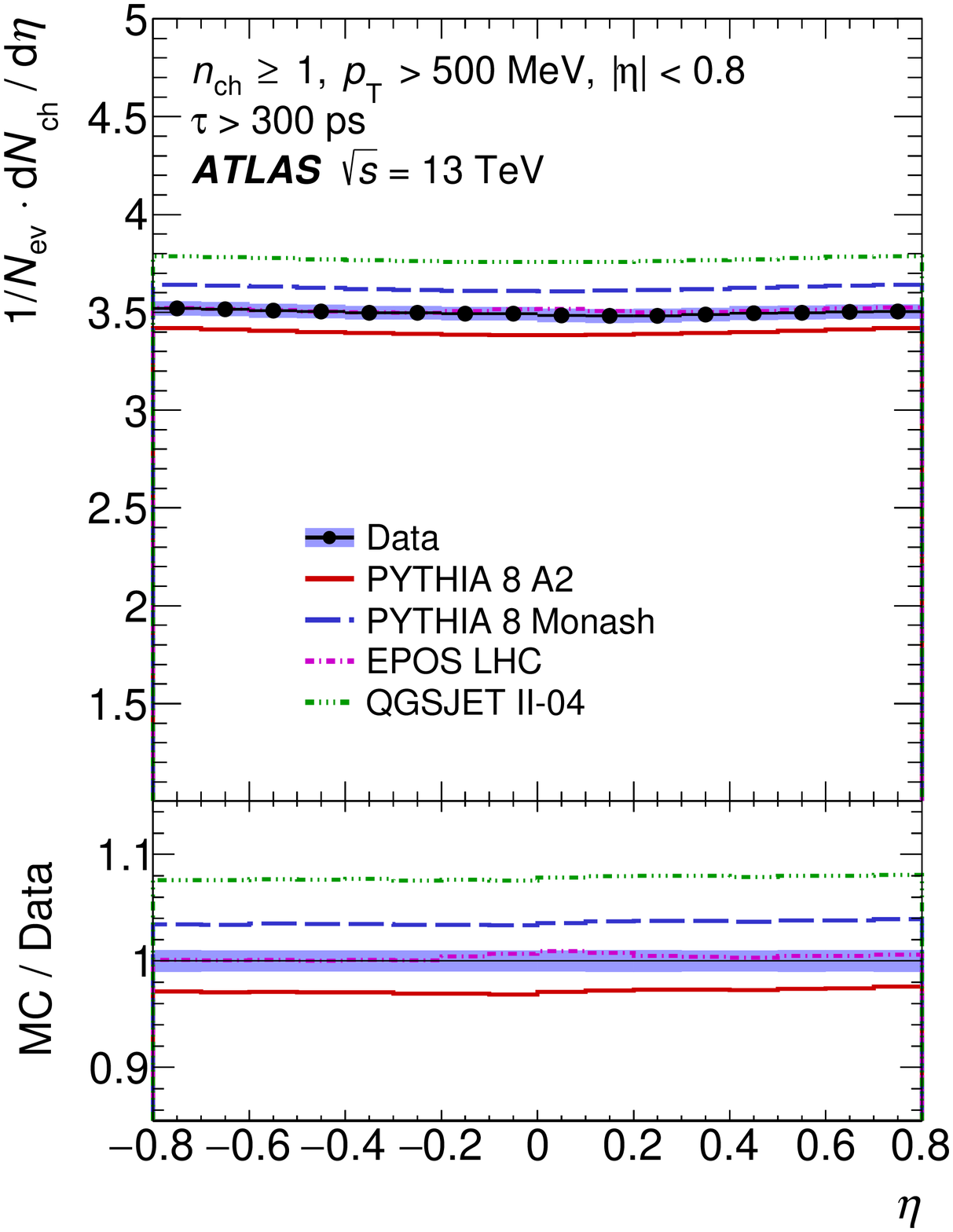}
                \label{fig:dndeta-reduced}
	}
	\subfigure[]{
		\includegraphics[width=0.43\textwidth]{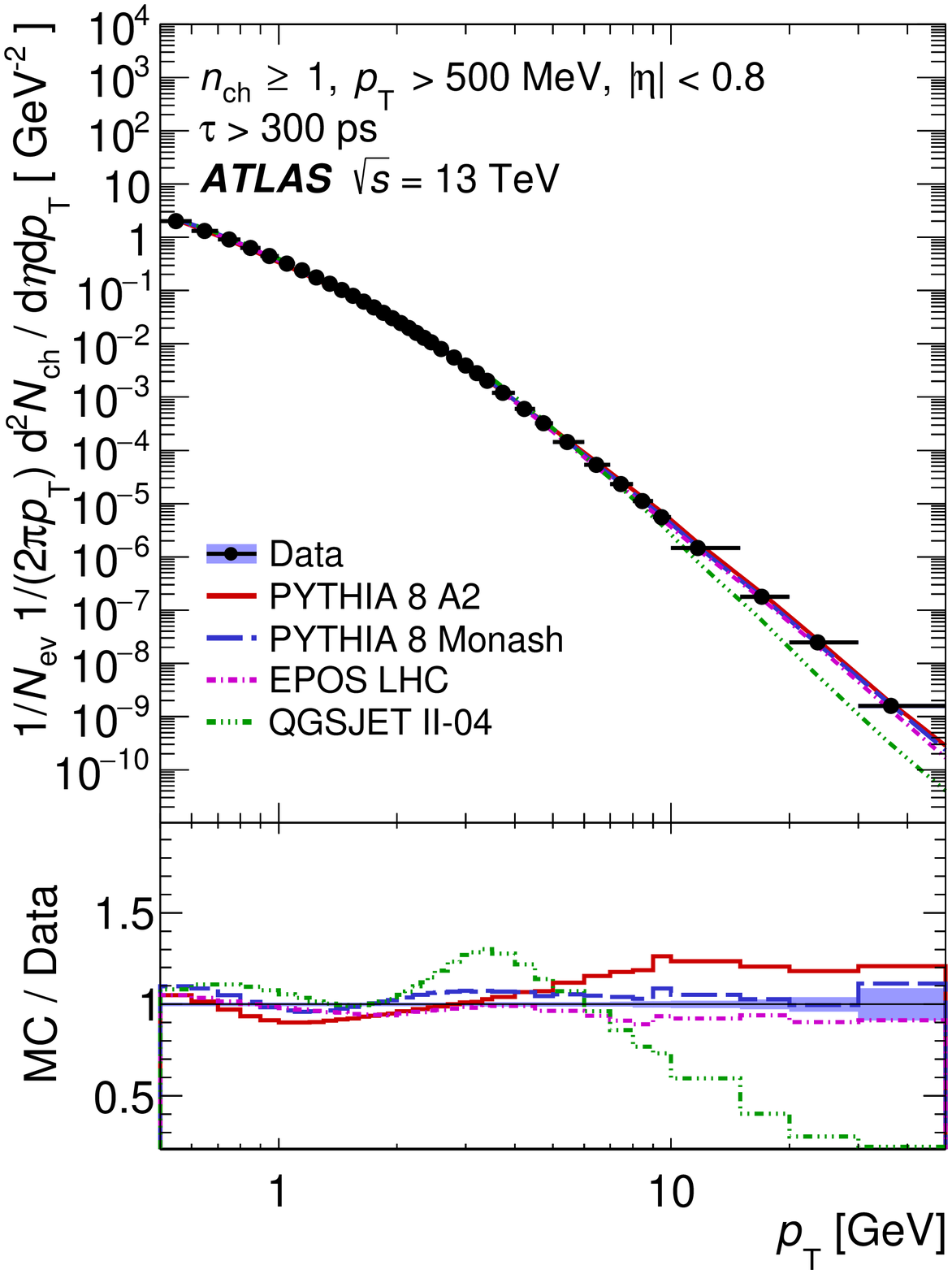}
                \label{fig:dndpt-reduced}
	}
	\subfigure[]{
		\includegraphics[width=0.43\textwidth]{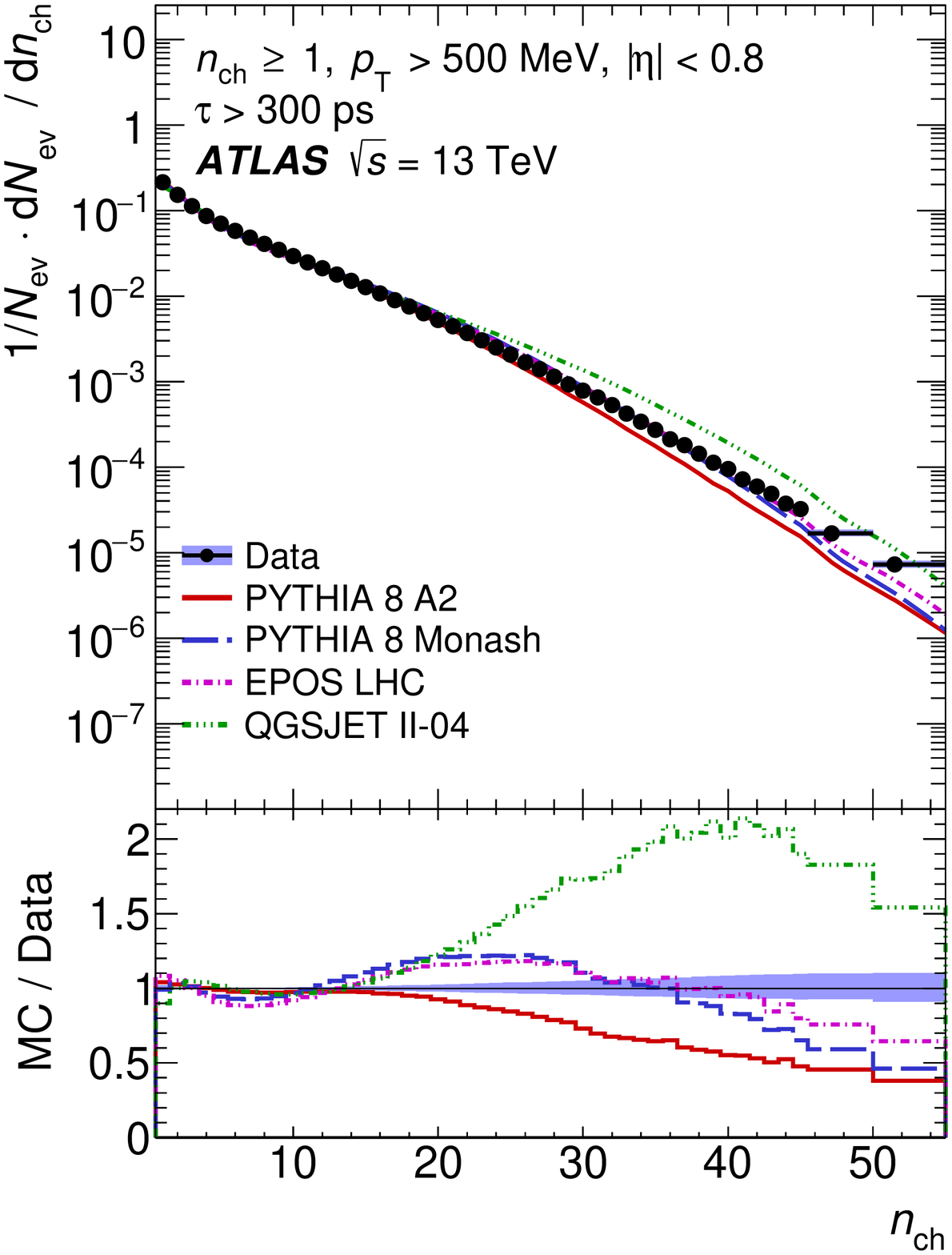}
                \label{fig:dNdnch-reduced}
	}
	\subfigure[]{
		\includegraphics[width=0.43\textwidth]{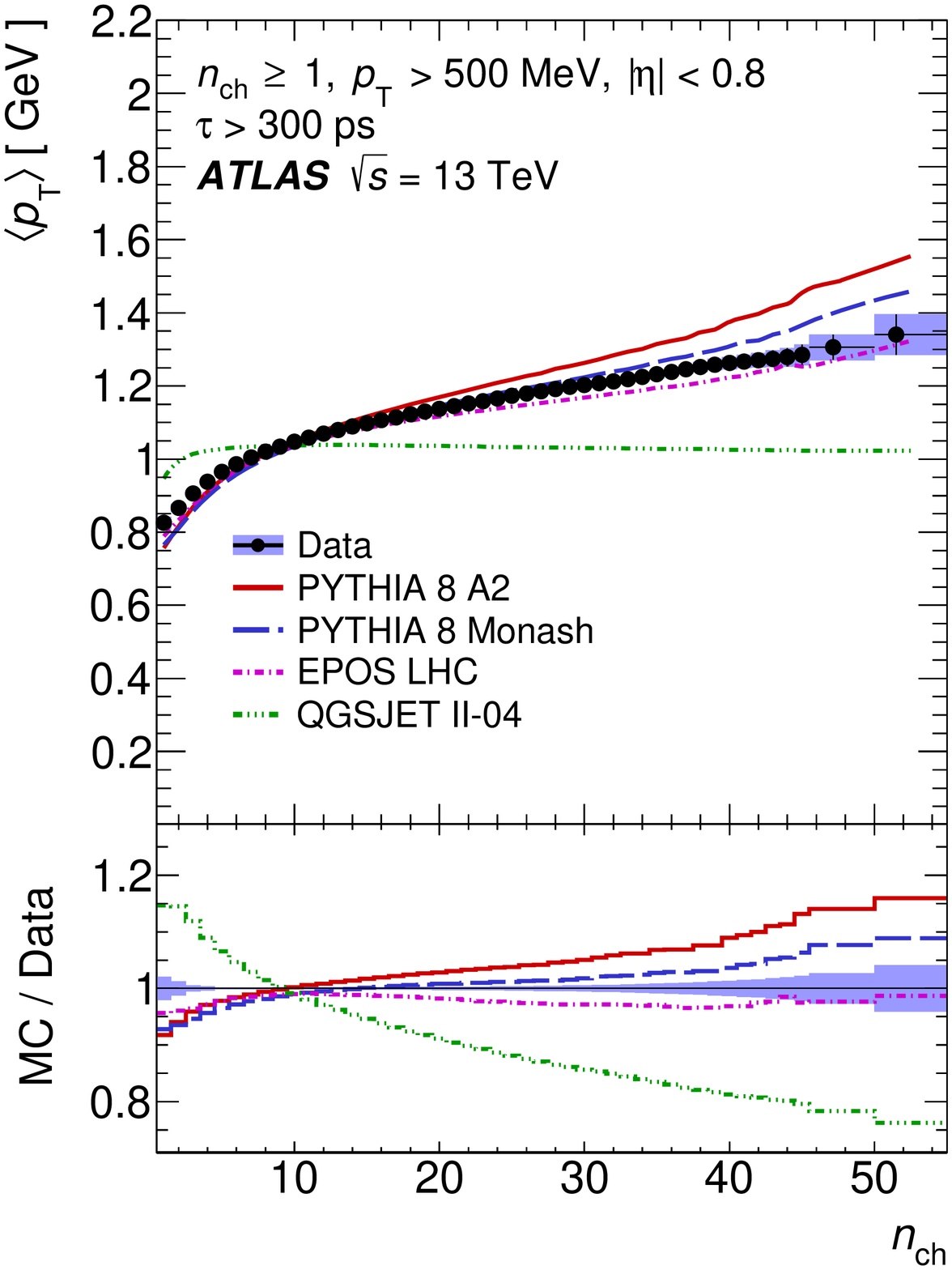}
                \label{fig:meanpt-reduced}
}
	\end{center}
	\caption {Primary-charged-particle multiplicities as a function of 
          (a) pseudorapidity, $\eta$, and (b) transverse momentum, \pta{}; (c) the
          multiplicity, \nch{}, distribution and (d) the mean transverse
          momentum, \meanpt{} , versus \nch{}
  in events with $\nch \geq 1$, $\pta > 500$~MeV and $|\eta| < 0.8$. 
The dots represent the data and the curves the predictions from different MC
models. The $x$-value in each bin corresponds to the bin centroid. The
vertical bars represent the statistical uncertainties, while the shaded
areas show statistical and systematic uncertainties added in quadrature. The
bottom panel in each figure shows the ratio of the MC simulation over the data. 
Since the bin centroid is different for data and simulation, the values of the
ratio correspond to the averages of the bin content.
}
  \label{fig:results-reduced}
\end{figure*}

The corrected distributions for primary charged particles in events
with $\nch \geq 1$ in the kinematic range $\pta > 500$~MeV and $|\eta|
< 0.8$ are shown in Figure~\ref{fig:results-reduced}. This is the phase space that is common to the ATLAS, CMS and ALICE experiments.

The method used to correct the distributions and obtain the systematic
uncertainties is exactly the same as
that used for the results with $|\eta|
< 2.5$, but obtained using the $|\eta|
< 0.8$ selection.

Figure~\ref{fig:dndeta-reduced} shows the primary-charged-particle multiplicity as a function of pseudorapidity, where the mean particle density is roughly 3.5, 
larger than in the main phase space due to the tighter restriction of
at least one primary charged particle with $|\eta| < 0.8$.
The \pta{} and \nch{} distributions are shown in
Figures~\ref{fig:dndpt-reduced} and~\ref{fig:dNdnch-reduced}
respectively and the \meanpt{} as a function of \nch{} is shown in
Figure~\ref{fig:meanpt-reduced}.
The level of agreement between the data and MC generator predictions follows the same pattern as seen in the main phase space.

\newpage 

\begin{flushleft}
{\Large The ATLAS Collaboration}

\bigskip

G.~Aad$^\textrm{\scriptsize 86}$,
B.~Abbott$^\textrm{\scriptsize 113}$,
J.~Abdallah$^\textrm{\scriptsize 151}$,
O.~Abdinov$^\textrm{\scriptsize 11}$,
B.~Abeloos$^\textrm{\scriptsize 117}$,
R.~Aben$^\textrm{\scriptsize 107}$,
M.~Abolins$^\textrm{\scriptsize 91}$,
O.S.~AbouZeid$^\textrm{\scriptsize 137}$,
N.L.~Abraham$^\textrm{\scriptsize 149}$,
H.~Abramowicz$^\textrm{\scriptsize 153}$
% ...
\bigskip
\\
% ...
$^{11}$ Institute of Physics, Azerbaijan Academy of Sciences, Baku, Azerbaijan\\
% ...
$^{86}$ CPPM, Aix-Marseille Universit{\'e} and CNRS/IN2P3, Marseille, France\\
% ...
$^{91}$ Department of Physics and Astronomy, Michigan State University, East Lansing MI, United States of America\\
% ...
$^{107}$ Nikhef National Institute for Subatomic Physics and University of Amsterdam, Amsterdam, Netherlands\\
% ...
$^{113}$ Homer L. Dodge Department of Physics and Astronomy, University of Oklahoma, Norman OK, United States of America\\
% ...
$^{117}$ LAL, Univ. Paris-Sud, CNRS/IN2P3, Universit{\'e} Paris-Saclay, Orsay, France\\
% ...
$^{137}$ Santa Cruz Institute for Particle Physics, University of California Santa Cruz, Santa Cruz CA, United States of America\\
% ...
$^{149}$ Department of Physics and Astronomy, University of Sussex, Brighton, United Kingdom\\
% ...
$^{151}$ Institute of Physics, Academia Sinica, Taipei, Taiwan\\
% ...
$^{153}$ Raymond and Beverly Sackler School of Physics and Astronomy, Tel Aviv University, Tel Aviv, Israel\\
% ...
$^{*}$ Deceased
\end{flushleft}


\end{document}